\documentclass[manuscript,screen]{acmart}
\usepackage{comment}
\usepackage{tabularx}
\usepackage{appendix}  
\usepackage{longtable}

\AtBeginDocument{%
  }

\setcopyright{acmlicensed}
\copyrightyear{2024}
\acmYear{2024}
\acmDOI{XXXXXXX.XXXXXXX}

\acmISBN{978-1-4503-XXXX-X/18/06}

\begin{document}

\title{Persona-L has Entered the Chat: Leveraging LLM and Ability-based Framework for Personas of People with Complex Needs} 

\author{Lipeipei Sun}
\affiliation{%
  \institution{Northeastern University}
  \city{Seattle}
  \state{WA}
  \country{USA}
} 
\email{sun.lipe@northeastern.edu}

\author{Tianzi Qin}
\authornote{These authors contributed to the first draft of the paper as well as the implementation of the user interface.}
\email{qin.tianz@northeastern.edu} 
\affiliation{%
  \institution{Northeastern University}
  \city{Vancouver}
  \state{BC}
  \country{Canada}
} 

\author{Anran Hu}
\authornotemark[1]
\email{hu.anr@northeastern.edu}
\affiliation{%
  \institution{Northeastern University}
  \city{Vancouver}
  \state{BC}
  \country{Canada}
} 

\author{Jiale Zhang}
\authornote{These authors contributed to the study and implementation of Large Language Modeling and Retrieval-Augmented Generation.}
\affiliation{%
  \institution{Northeastern University}
  \city{Vancouver}
  \state{BC}
  \country{Canada}
} 
\email{zhang.jiale2@northeastern.edu}

\author{Shuojia Lin}
\authornotemark[2]
\affiliation{%
  \institution{Northeastern University}
  \city{Vancouver}
  \state{BC}
  \country{Canada}
} 
\email{lin.shuo@northeastern.edu}

\author{Jianyan Chen}
\authornotemark[2]
\affiliation{%
  \institution{Northeastern University}
  \city{Vancouver}
  \state{BC}
  \country{Canada}
} 
\email{chen.jianya@northeastern.edu}

\author{Mona Ali}
\affiliation{%
  \institution{Northeastern University}
  \city{Vancouver}
  \state{BC}
  \country{Canada}
} 
\email{mon.ali@northeastern.edu}

\author{Mirjana Prpa}
\affiliation{%
  \institution{Northeastern University}
  \city{Vancouver}
  \state{BC}
  \country{Canada}
} 
\email{m.prpa@northeastern.edu}

\keywords{Persona, UX Design, Context, Ability-based Framework}


\maketitle

\begin{figure}[h]
    \centering
    \includegraphics[width=1\linewidth]{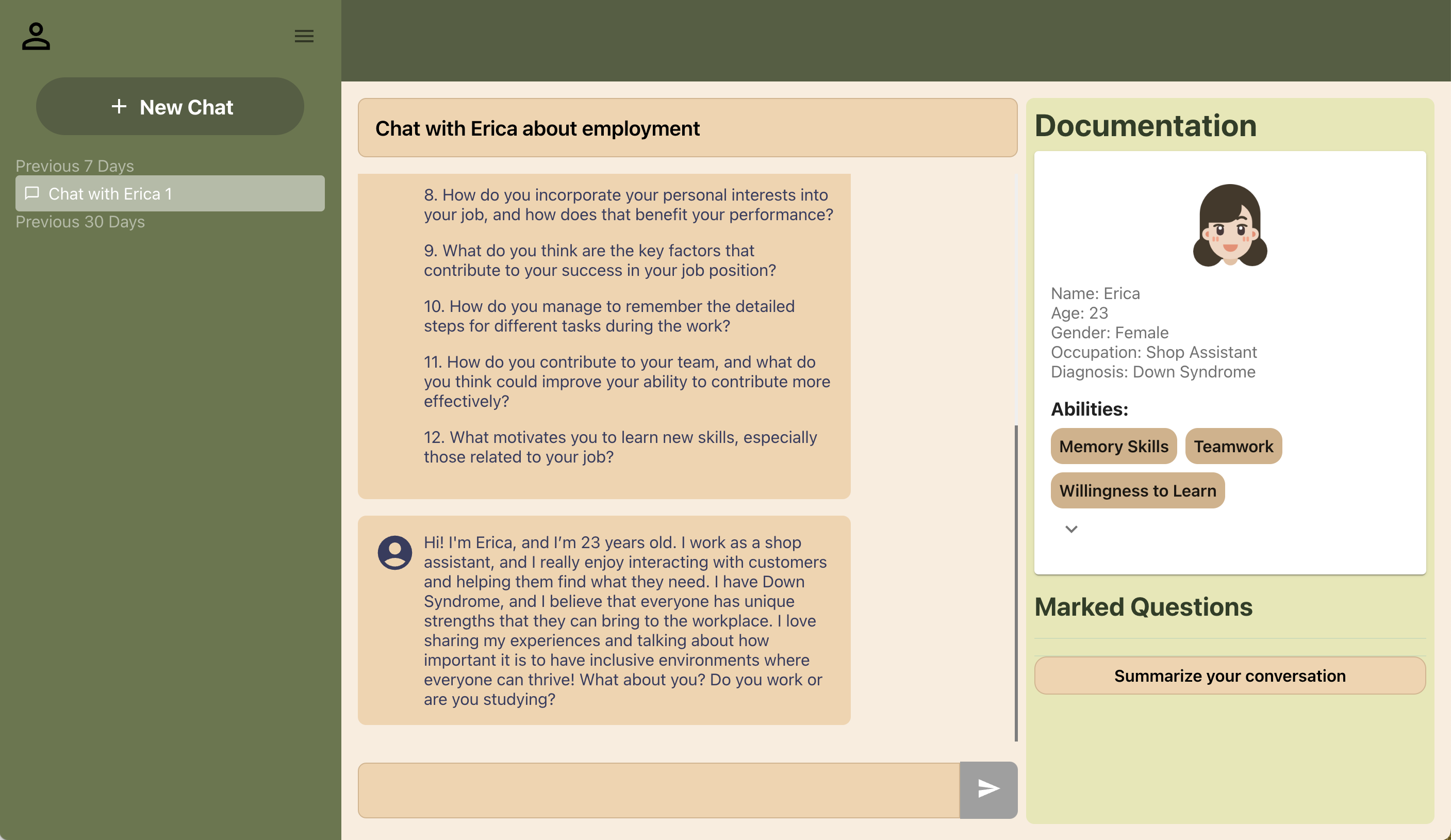}
    \caption{Persona-L Interface  featuring a chat window where users can interact with created Persona-L: Erica about previously selected Theme: Employment.}
    \Description{This figure shows the Persona-L interface, featuring a chat window where users can interact with Persona-L through conversation.}
    \label{fig:Persona-L}
\end{figure}

\section{Abstract}

We present \textit{Persona-L}, a novel approach for creating personas using Large Language Models (LLMs) and an ability-based framework, specifically designed to improve the representation of users with complex needs. Traditional methods of persona creation often fall short of accurately depicting the dynamic and diverse nature of complex needs, resulting in oversimplified or stereotypical profiles. \textit{Persona-L} enables users to create and interact with personas through a chat interface. \textit{Persona-L} was evaluated through interviews with UX designers (N=6), where we examined its effectiveness in reflecting the complexities of lived experiences of people with complex needs. We report our findings that indicate the potential of Persona-L to increase empathy and understanding of complex needs while also revealing the  need for transparency of data used in persona creation, the role of the language and tone, and the need to provide a more balanced presentation of abilities with constraints.

\section{Author Keywords}
Persona; Human-Computer Interaction (HCI); UX Design; Large-Language Models (LLMs); Context; Ability-based Framework

\section{Introduction}
Personas are detailed and archetypal user profiles that represent a holistic image of a segment of the target audience by providing key insights about the user's goals, interests, and behaviors \cite{grudin2006personas}. As such, personas are used in various domains, including research, software development, healthcare and marketing \cite{salminen_use_2022} as one of the available tools to help guide the understanding of user needs and behaviors which has been shown to enhance the usability and effectiveness of designs \cite{nielsen2013personas}. Personas are also widely taught and leveraged tool in educational settings \cite{barambones2024chatgpt,goel2023preparing}, where access to users and conducting primary research is often limited \cite{barambones2024chatgpt}.

Persona creation has traditionally relied on data from qualitative methods such as interviews, observations, and survey data of target users. More recent methods use large datasets, including statistical analysis of customer data \cite{mcginn_data-driven_2008}, clickstreams \cite{zhang_data-driven_2016}, and social media data \cite{jung_persona_2017}. However, these approaches often do not yield optimal results when empirical data is restricted due to practical and ethical reasons, and when large datasets are not readily available. In addition, in the data-driven persona creation process, designers do their best to use available data, but the data sources might not always be \textbf{trustworthy} \cite{salminen_ethics_2020}.
 Most recently, large language models (LLMs), like OpenAI's GPT-3.5 and GPT-4, have been used to generate synthetic research data and demonstrated promising capabilities in generating text that is both coherent and contextually appropriate \cite{hamalainen2023evaluating}, that could potentially be leveraged in the creation of personas. Recent studies also indicate that LLMs could significantly reduce the time and effort required to generate personas, as they can quickly produce detailed and diverse personas based on large datasets \cite{hamalainen2023evaluating}.
However, lack of data and limited access to users in primary research impacts the quality and depth of insights in created personas, resulting in personas that are often \textbf{abstract, impersonal, and misleading}\cite{matthews2012designers}. This challenge can be potentially amplified by LLMs in LLM-generated personas that represent user groups whose data was not included in  LLM training, which may lead to oversimplification and stereotyping in created personas \cite{barambones2024chatgpt}. This is particularly critical to address when creating personas of people with complex needs (including people with developmental, cognitive, or physical disabilities) \cite{edwards2020three}. When personas are used to depict users with complex needs, generalized user archetypes can result in more simplistic or stereotyped representations \cite{marsden2016stereotypes} and fail to capture the complexity of lived and real-world experiences \cite{theil_23}.  

In this paper, we present our initial efforts towards a larger research initiative focused on investigating opportunities and limitations of using LLMs in persona creation for diverse groups of users. In particular, we present Persona-L (see Figure \ref{fig:Persona-L}), a web-based tool for the creation of personas of people with complex needs. We chose modeling personas of people with  Down Syndrome as the use case for this research. To ground LLM responses we leverage manually curated data collected from the publicly available forums on Down Syndrome to train LLM (GPT-4o mini). We use the LLM and Retrieval Augmented Generation (RAG) framework to enhance the correctness of LLM responses and ground our design in an ability-based framework to provide more comprehensive insights into the abilities and disabilities of people with Down Syndrome. This approach offers a multifaceted representation of complex needs by providing descriptions of abilities (rather than just disabilities) for painting a more comprehensive context that has been found to help reduce the negative stereotypes towards people with complex needs \cite{nolte2022creating}.

To that end, our research is focused on answering the following research question (RQ): How can we leverage LLMs and an ability-based framework to build personas of people with complex needs? In answering our RQ we made the following contributions: (1) Development of Persona-L, an interactive web-based tool that leverages LLM (GPT-4o mini) and Retrieval-Augmented Generation (RAG) to create personas of people with complex needs, (2) Data collection and integration of an ability-based framework for people with Down Syndrome in Persona-L, (3) Evaluation of  Persona-L interface and generated personas. Our proposed solution is designed to foster a deeper and more empathetic understanding of the unique needs and behaviors of people with complex needs. Our work contributes to the discourse on UX design methods, in particular the use of LLMs for persona creation and personas of people with complex needs. In this paper, we present the case study of leveraging Persona-L for creating personas of people with Down Syndrome due to our dataset containing accounts of lived experiences of people with Down syndrome related to 3 themes: Education, Employment, and Family. Yet, our approach is applicable to other user groups and themes for which the data is available.
\section{Related Work}
\subsection{Approaches for Creating Personas}
Persona are composite archetypes created from user behavior patterns identified during research \cite{cooper2014about}. Personas should be developed with rigor and sophistication, avoiding stereotypes and ensuring they accurately represent a meaningful cross-section of users. In addition to the creation of personas that rely on qualitative data gathered in interviews and ethnography, new approaches for designing personas emerged including data-driven personas \cite{jansen2022create}, and most recently LLM generated personas.

\subsubsection{Data-driven Personas}
Data-driven personas rely on large datasets including statistical analysis of customer data \cite{mcginn_data-driven_2008}, clickstreams \cite{zhang_data-driven_2016}, and social media data \cite{jung_persona_2017}. The use of real-time online user data has been explored in data-driven personas in efforts to keep personas up-to-date  \cite{jung_persona_2017}. Previous research has found data-driven personas creation to be efficient, cost-effective, scalable, and reflective of actual user behavior \cite{zhang_data-driven_2016,mcginn_data-driven_2008}
However, it also has been found that data-driven personas may be susceptible to biases. It has been proposed that generating more personas results in a better representation of diverse demographic groups that can help reduce biases in them  \cite{salminen_detecting_2019, salminen_creating_2022}. 

\subsubsection{LLM -Generated Personas}
With the rapid development of LLMs and their capacity to generate human-like responses \cite{tavast2022language}, a growing body of research has highlighted the potential for using LLMs in various aspects of creating believable and relatable personas \cite{salminen_deus_2024,zhang_auto-generated_2024}. For example, \cite{goel2023preparing, schuller2024generating} have demonstrated the use of prompt strategies for persona creation.  \cite{barambones2024chatgpt} used ChatGPT to simulate various personas in qualitative interviews. \cite{depaoli2024improved} employed LLMs to facilitate thematic analysis of qualitative interviews and accurately capture user behaviors and personality traits, thereby enhancing the realism of created personas \cite{depaoli2024improved}. 

It has also been shown that LLM-generated personas, compared to traditional methods, resulted in improved diversity and accuracy \cite{zhang_auto-generated_2024}. LLMs can generate consistent, believable, relatable, and informative personas while offering a more efficient and flexible approach to persona creation than traditional methods making it less resource-intensive\cite{depaoli2024improved, schuller2024generating}. Previous research also highlighted the potential for the use of LLMs in the creation of effective personas for educational purposes \cite{goel2023preparing, barambones2024chatgpt}. Although LLMs have shown great promise, many LLM-related drawbacks impact the creation and application of personas. \cite{salminen_deus_2024} showed that biases related to age, occupation, and geographic representation can persist in LLM-generated personas, and it was suggested that careful attention must be paid to addressing biases and ensuring the authenticity of LLM-generated personas through iterative validation and engagement with users and subject experts.  

Another limitation identified is that the LLM-generated personas can be too generic, due to lack of specificity of user groups in LLM training data. \cite{goel2023preparing} found LLMs-generated personas to be rated lower by UX professionals in terms of willingness to use, credibility, empathy, and memorability, which is attributed to generic responses and inconsistencies after multiple prompting \cite{goel2023preparing}. \cite{schuller2024generating} highlighted issues with stereotypes and the lack of empathy. \cite{barambones2024chatgpt} explored prompting strategies to address limitations in LLM-generated responses such as repetition, low variability, and gender bias however this is still under-explored area.

\subsection{Challenges and Limitations in Personas} 
Personas are widely used in the product design phase to help guide a better understanding of user needs in ideation, decision-making, and communication of design work. However, over time some concerns have been raised about the use of Personas. \cite{matthews2012designers} discussed challenges in personas being \textbf{abstract, impersonal, and misleading}. \cite{oudshoorn2004configuring} pointed out that a persona (a representative of a user group) usually has characteristics such as age, gender, and race, but in this representation process, diversity inevitably decreases due to the \textbf{limited number of personas}. \cite{marsden2016stereotypes} noted that personas focus on typical qualities, which may lead to \textbf{misalignment with the multiple identities of user groups}, causing a false sense of understanding. \cite{marsden2016stereotypes} discussed the biases involved in creating personas, noting that personas can be superficial and activate stereotypes, and suggested that the multiple identities of user groups should be considered in the process of creating personas. In addition, \cite{edwards2020three} suggested that traditional methods of creating personas frequently fail to capture the full\textbf{ contextual depth—including social, cultural, and personal factors—that significantly influences how target users interact with technology}.

\subsection{Developing Personas representing People with Complex Needs} 
Creating personas that represent the identities and needs of people with complex needs may require additional considerations compared to personas more aligned with "general" populations. \cite{edwards2020three} presented the challenges in creating personas of populations with complex needs, particularly those with disabilities and intersecting identities. This research highlights that personas are typically static, while the identities of people with disabilities are dynamic and ever-evolving. Additionally, the number of personas used in any project is inherently limited, which can fail to represent the diversity and complexity of lived experiences effectively.  Furthermore, because personas are fixed documents, they cannot adapt or respond as a real person would during interactions, potentially perpetuating stereotypes \cite{edwards2020three}. 

Several efforts have explored how to create personas representing people with complex needs. \cite{neate2019co} describes a study that explores the use of co-created personas as a method in the co-design process with users who have diverse needs, specifically those with Parkinson’s disease, dementia, or aphasia. \cite{schulz2012creating} advises on how to better include people with disabilities in the persona creation process to build reliable personas. In a recent study, \cite{nolte2022creating} presents an \textbf{ability-based approach} to create personas for Deaf and Hard of Hearing (DHH) individuals within the travel context. Traditional persona methods often misrepresent DHH individuals by labeling them through their impairments. \cite{nolte2022creating}  argue for an ability-based approach that aligns with how DHH individuals view themselves. In addition, Nolte et al. emphasize the importance of shifting from traditional user modeling techniques that often focus on disabilities, advocating instead for \textit{a focus on the user's abilities, contextualizing these abilities within specific tasks and environments} and provides two use cases applying this approach (Use Case I: Communication for Weaning Patients and Use Case II: Travel Information for Sign Language Users)\cite{nolte2022implementing,nolte2022creating}

\subsection{Implications for Interactive and Dynamic User Interface for LLM-generated Personas}  
In addition to a body of research that explored the efficacy of using LLMs in persona generation, previous research provided insights about how to design a dynamic and interactive user interface for LLM personas to minimize drawbacks present in persona creation, such as providing wider context to persona responses and approaches to amplify empathy towards user groups that LLM-personas represent. 
\cite{ha2024clochat} proposed prioritizing persona customization, reducing the initial setup burden for LLM interfaces, and implementing adaptive algorithms that tailor persona behaviors based on user intentions, enabling LLM- personas to evolve, significantly enhancing the capability of eliciting empathy towards the persona. 
\cite{cheng2023compost} explored the use of specific topics and multifaceted personas in LLM simulations, emphasizing the importance of documenting the creators' positions, motivations, and processes. This transparency is essential since it helps understand the inherent biases that could affect the final LLM persona outputs. \cite{subramonyam2024bridging} introduced the \textit{Transactional Model of Communication}, which facilitates \textbf{iterative feedback and adjustments within LLM interactions,} making it suitable for interfaces like ChatGPT, \textbf{where users can edit previous prompts to refine the interaction, therefore shaping a more dynamic and rich context for persona responses}. Lastly, \cite{schmidt2024simulating} advocates for the creation of dynamic personas that \textbf{utilize prompts to highlight the main characteristics or needs of the persona}. They argue that this method allows for iterative refinement of personas, making them more accurate and reflective of the users' evolving needs and behaviors instead of the static, generic ones.

\subsection{Positioning of Our Research in the Context of Related Work} Previous studies highlighted issues with personas such as being abstract, impersonal, and misleading \cite{matthews2012designers},
lacking contextual depth —including social, cultural, and personal factors—that significantly influences how target users interact with technology \cite{edwards2020three}, and limited in presenting multiple identities of user groups \cite{marsden2016stereotypes} due to fairly limited number personas that are usually created \cite{oudshoorn2004configuring}, potentially susceptible of perpetuating stereotypes, especially when creating personas of people with complex needs \cite{edwards2020three}.

Our research contributes to the existing literature by focusing specifically on the development of an interactive LLMs-persona interface for the creation of personas of people with complex needs (in this research we focus on a particular use case of modeling personas for people with Down Syndrome) grounded in the ability-based framework. This approach not only could potentially address the gaps identified in current LLM persona creation but also marks the first implementation of an ability-based framework within LLM applications to provide insights into multifaceted experiences when creating personas of people with complex needs. Unlike traditional LLM applications that rely heavily on pre-defined templates resulting in generic and stereotypical representations, our research emphasizes real-time interaction and persona responses grounded in data shared by people with Down Syndrome. Our approach allows LLMs to dynamically produce personas that are not only contextually richer but also potentially reflect the actual capabilities and self-identities of people with complex needs, thereby potentially reducing stereotypes and enhancing empathetic engagement. To this end, in our system, we leverage RAGs and data collected from publicly available forums on Down Syndrome (Canadian Down Syndrome Society (CDSS) \cite{CDSS}, WorldDownSyndromeDay \cite{WDSD}) to augment LLM output and ensure the correctness of the LLM responses while enabling users to inquire about the wider context related to the topics of interest.

\section{Persona-L System Design}   
Persona-L system design is shown in Figure \ref{fig:System Design Diagram}. Detailed methodology is described in Appendix \ref{Methodology} and involved data collection, theme categorization and formulation of ground truth questions, training LLM model on collected data, and integration of the back-end LLM model with the user interface to request data through APIs, making sure the system meets its intended usability and interaction goals. In the following sections, we describe the key aspects of Persona-L.
\begin{figure}
    \centering
    \includegraphics[width=1\linewidth]{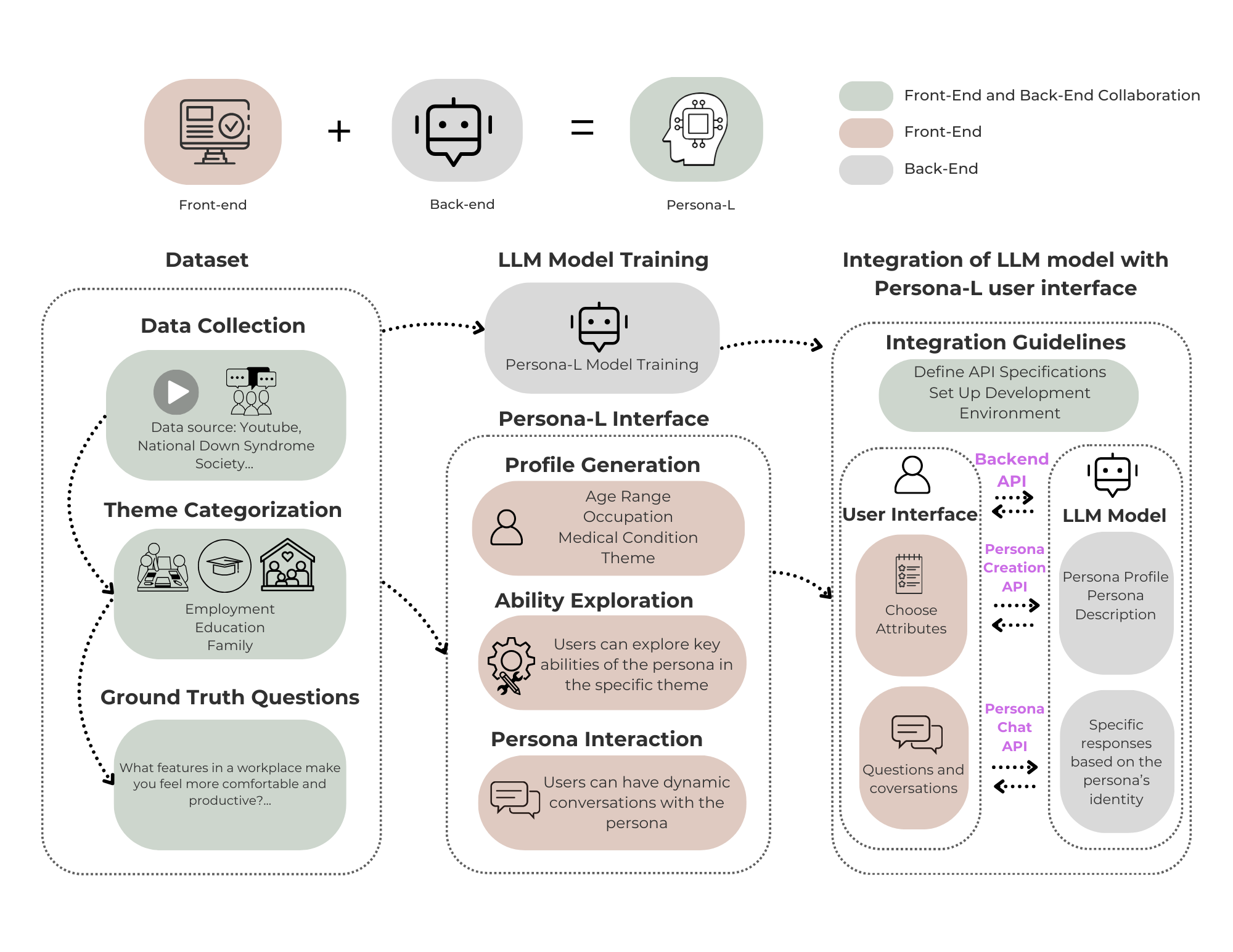}
    \caption{System Design Diagram. It maps the journey from data collection to interactive persona creation, highlighting the collaborative efforts required to deliver a sophisticated and engaging user experience.}
    \label{fig:System Design Diagram}
\end{figure}

\subsection{An Ability-Based Framework for LLM persona}
The concept of Ability-Based Design, introduced by Wobbrock et al. \cite{wobbrock2011ability, wobbrock2018ability}, addresses the challenge of creating accessible technologies and has been a topic of discussion in human-computer interaction research for over a decade. This approach shifts the focus from typical user modeling that often emphasizes disabilities to one that highlights their abilities. It suggests significant modifications to conventional user modeling techniques, which are often inadequate for fully capturing the abilities of individuals \cite{nolte2022implementing}. In creating personas for people with complex needs, many existing approaches focus predominantly on disabilities, leading to a misalignment with self-perception and often resulting in the needs of these groups being overlooked or insufficiently represented. This misrepresentation can generate tensions and challenges \cite{nolte2022creating, edwards2020three}, which is the issue we aim to address in our research.

To this end, during the data mining phase, we pay special attention to collecting data about the abilities and strengths of people with complex needs. In the interface design phase, we highlight the ability traits of personas. For example, we introduce the capabilities of individuals with Down syndrome in the interface, provide options for exploring these abilities, and allow users to select them. This enables the creation of personas with diverse abilities, that inform responses of personas across different thematical contexts, thus fostering a more inclusive and representative interaction model.

Through such an Ability-Based Approach, we aim to achieve two key outcomes:
\begin{itemize}
    \item \textbf{Promoting alignment with the self-identities of people with complex needs: }People with complex needs do not want to be identified as disabled or problematic \cite{nolte2022creating}. Our approach ensures that the personas we create are more aligned with how individuals see themselves and how they wish to be represented. This can, we hypothesize, better reflect the identities of people with complex needs.
    \item \textbf{Enhancing empathy and reducing stereotypes:} Currently, the lack of attention to people with complex needs and the generic methods used to create LLM-based personas often result in perpetuation of biases and stereotypes \cite{barambones2024chatgpt,marsden2016stereotypes}. We aim that by grounding persona responses in data shared by people with complex needs, which fully showcases their competence and potential, we can overcome or at least, reduce stereotypes. This not only improves the accuracy and representation of the personas but also plays a crucial role in societal perceptions, encouraging a more inclusive view of diversity. 
\end{itemize}

\subsection{User Interface Design}
\begin{figure}
    \centering
    \includegraphics[width=1.1\linewidth]{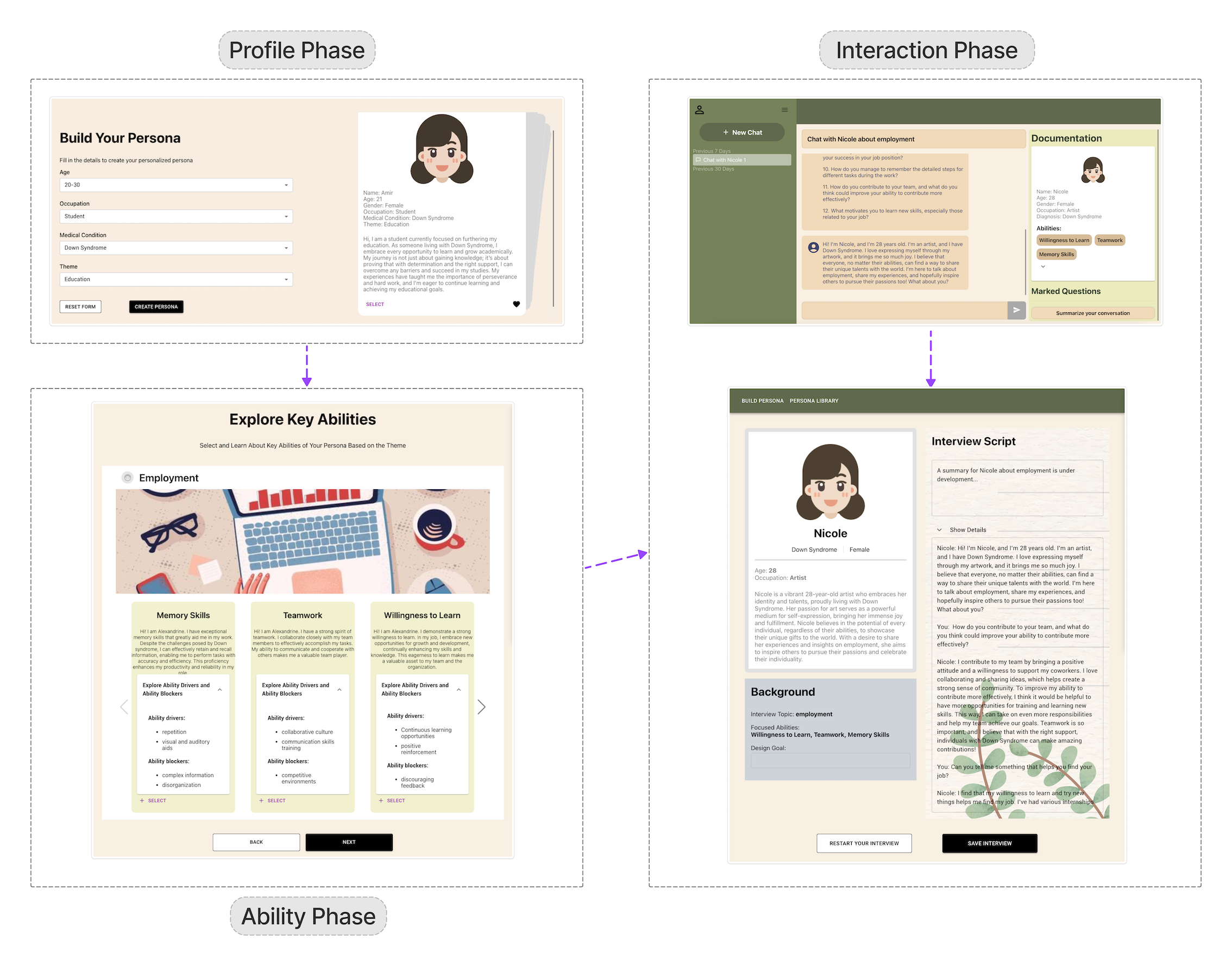}
    \caption{Phases in User Interface Design: The "Profile Phase" involves persona customization and theme selection. The "Ability Phase" allows users to explore and select key abilities under a specific theme. The "Interaction Phase" facilitates user engagement with the personas LLMs and provides tools for saving and documenting interactions. }
    \label{fig:Phases}
\end{figure}

Our system design addresses the limitations of current LLM-created personas, which are often overly simplistic and lack context that can provide deeper insights. By leveraging LLMs to simulate personas rather than generate basic descriptions of personas, we aim to create more nuanced and empathetic character profiles. This approach will enhance the depth and authenticity of the personas, providing a more natural and believable representation of the user group\cite{shao2023character}. 

Using the "4W1H" guideline for designing a human-LLM interaction system, we can clearly define the structure and objectives of our system \cite{gao2024taxonomy}. \textbf{Who:} Our intended end users are UX designers, researchers, developers, and other stakeholders who interact with the system to create personas of people with complex needs. In this proof-of-concept, we focus on personas representing people with Down syndrome. \textbf{What:} The objective is to enable users to create personas and chat with them, allowing users to gather richer context around the topics of interest. \textbf{Where:} The interface is specifically designed for the web. \textbf{When:} The system is structured into three primary phases: Profile Phase, Theme/Ability Phase, and Interaction Phase (see Figure \ref{fig:Phases}). In the Profile Phase, users create an initial persona. During the Theme/Ability and Interaction Phases, users engage with LLM-simulated personas to enhance their understanding of the persona within a specific context (theme). Once created personas can be revisited multiple times and used when needed. For example, one may create a persona and engage with the persona throughout an ideation phase, and come back to it later during decision-making. \textbf{How:} Leveraging the computational power of GPT-4o mini and an ability-based framework for people with complex needs, our system focuses on enabling the creation of detailed, dynamic, and adaptable personas that genuinely reflect the diverse abilities and needs of the people with Down syndrome. 

We describe the capabilities of our system through three Phases of interaction and by providing user stories to paint the picture of how we envision users to interact with Persona-L. For a detailed description of Persona-L features please refer to the Table \ref{tab:features} in Appendix \ref{phase3 user stories}.

\subsubsection{\textbf{Phase 1: Profile Phase}} The objective is to enable users to develop initial personas, setting the stage for subsequent interactions.
\begin{figure}
    \centering
    \includegraphics[width=1\linewidth]{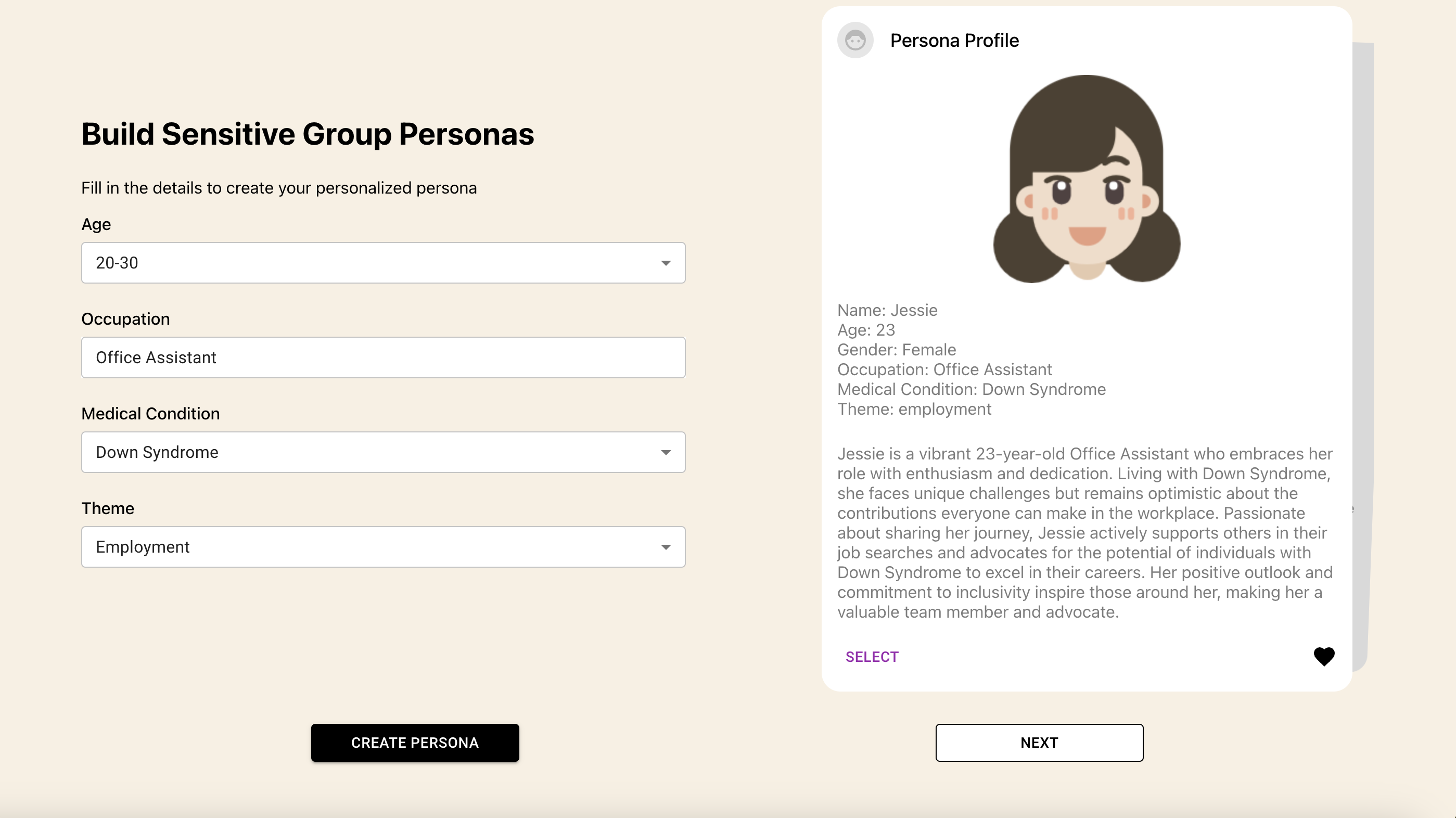}
    \caption{Profile Phase Page}
    \label{fig:Profile Phase Page}
\end{figure}

\textit{Vignette}: Jane, a junior UX Designer, is tasked to explore the challenges that young professionals with Down Syndrome face when looking for jobs, create a persona, and ideate a couple of design ideas that may solve those challenges. Jane is intimidated by the task since she has never had a close interaction with people with Down Syndrome and doesn't know where to start. Jane needs a tool that can help her learn about this particular user group, specifically in the context of employment (theme), before conducting in-person interviews. \\Persona-L supports this objective with the following features (see Figure \ref{fig:Profile Phase Page}):

 \begin{itemize}
    \item \textbf{Persona attribute Customization:} Multiple identities influence a person's self-perception and experiences. This approach may help to minimize stereotypical attributions and promote a more ethical and responsible representation of users \cite{marsden2019personas}. Users can personalize the initial settings by selecting from a set of predefined or custom demographic and social attributes to create a concise profile of the target persona ( such as age and occupation). 
    \item \textbf{Theme Selection:} Users select themes that will influence the persona's interaction scenarios. We focus on 3 themes: employment, education, and family for people with Down Syndrome, for which we collected data. 
Setting themes is consistent with the latest emerging practices: Specify different aspects and then use an LLM to create elaborate personas and scenarios, which lead to mitigating caricature\cite{schmidt2024simulating, cheng2023compost}. Moreover, theme-led conversations significantly increase the perceived anthropomorphism of the chatbot and enhance the pleasurable aspects of interaction\cite{haugeland2022understanding}. \item \textbf{Persona Customization:} After selecting initial attributes, a persona based on those attributes will be created, allowing users to save and revisit personas as needed. 
 \end{itemize}

\subsubsection{\textbf{Phase 2: Theme/Ability Phase}} The objective is to facilitate deeper engagement with key abilities in the specific themes relevant to the persona.
    
\begin{figure}
    \centering
    \includegraphics[width=0.75\linewidth]{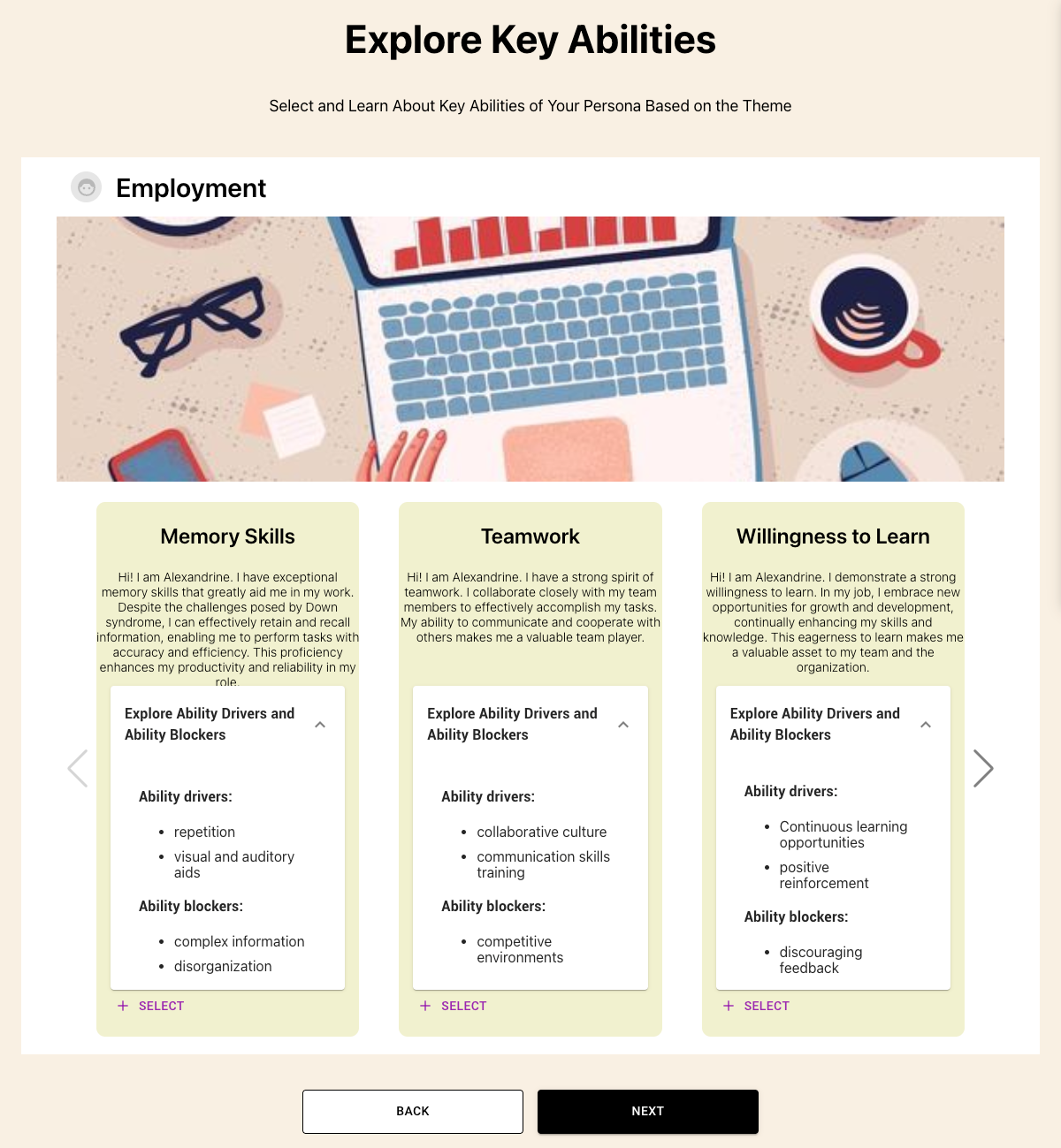}
    \caption{Persona-L: Display of Abilities per selected Theme:Employment }
    \label{fig:ability phase}
\end{figure}

\textit{Vignette:} Jane wants to know what abilities people with Down Syndrome have that can support their Employment (theme). In particular, she would like to learn about the factors that enhance (ability drivers) and hinder (ability blockers) the abilities of individuals with Down Syndrome regarding their employment. By focusing on abilities, rather than just disabilities, Jane can enrich the persona profiles with more context and potentially minimize any stereotypes she may hold. \\Persona-L supports this objective with the following features (see Figure \ref{fig:ability phase}):

\begin{itemize}
    \item \textbf{Abilities Contextualized:} Depending on the theme selected by the user, different key abilities will be displayed. Each ability will be briefly and informatively introduced, reflecting the real-life narratives of individuals with Down syndrome. An ability model emphasizes what users can do rather than what they cannot; beyond demographic information\cite{nolte2022implementing, wobbrock2018ability}, it could include more details and extend the context of the persona to address the special needs of the people with complex needs. As advocated by \cite{nolte2022implementing,nolte2022creating}, modeling of persona benefits from contextualizing abilities within specific themes (task and environments).
    \item \textbf{Ability Enhancers and Blockers:} Enabling and disabling factors for each ability will be displayed. For each ability driver and blocker, a short story will appear when the mouse hovers over them, helping users better understand the role these abilities play in the daily lives of individuals with Down syndrome. Users can involve several abilities in different themes to see potential impacts for the following interactive persona crafting (details provided in Table ~\ref{tab:ability}). Rather than considering user abilities in isolation, this approach examines how these abilities interact with specific contexts in which some tasks are performed. This helps in identifying which abilities are relevant for each task and how they might be affected by different contexts\cite{nolte2022implementing}.             
\end{itemize}

\begin{table}[ht]
    \centering
    \begin{tabular}{|c|p{0.6\textwidth}|l|} \hline
    
         Theme&  Employment\\ \hline 
         Ability Name&  Memory Skills\\ \hline 
 Ability Description&Hi! I have exceptional memory skills that greatly aid me in my work. Despite the challenges posed by Down syndrome, I can effectively retain and recall information, enabling me to perform tasks with accuracy and efficiency. This proficiency enhances my productivity and reliability in my role.\\ \hline 
 Ability Driver&Visual and Auditory aids\\ \hline 
 Ability Driver Story&In my workspace, I use visual aids like color-coded labels and auditory aids like recorded instructions. For instance, I have labels with pictures on the storage boxes and listen to instructions on my device. These aids provide strong cues that help me remember important details and perform my tasks accurately.\\ \hline 
 Ability Blocker&Complex Information\\ \hline 
 Ability Blocker Story&When I encounter complex information, it can be challenging for me to process and retain it. I remember a time when I had to learn a detailed procedure for a new task. It was overwhelming until my supervisor broke it down into simpler steps, which made it easier for me to understand and remember.\\ \hline
    \end{tabular}
    \caption{One example of ability name, ability driver and ability blocker under employment theme }
    \label{tab:ability}
\end{table}

\subsubsection{\textbf{Phase 3: Interaction Phase - Conversational System}} The objective is to enable users to engage in a conversation with the persona they developed and inquire about the topics they are interested in learning more about the persona's experiences and pain points.

\begin{figure}
    \centering
    \includegraphics[width=0.7\linewidth]{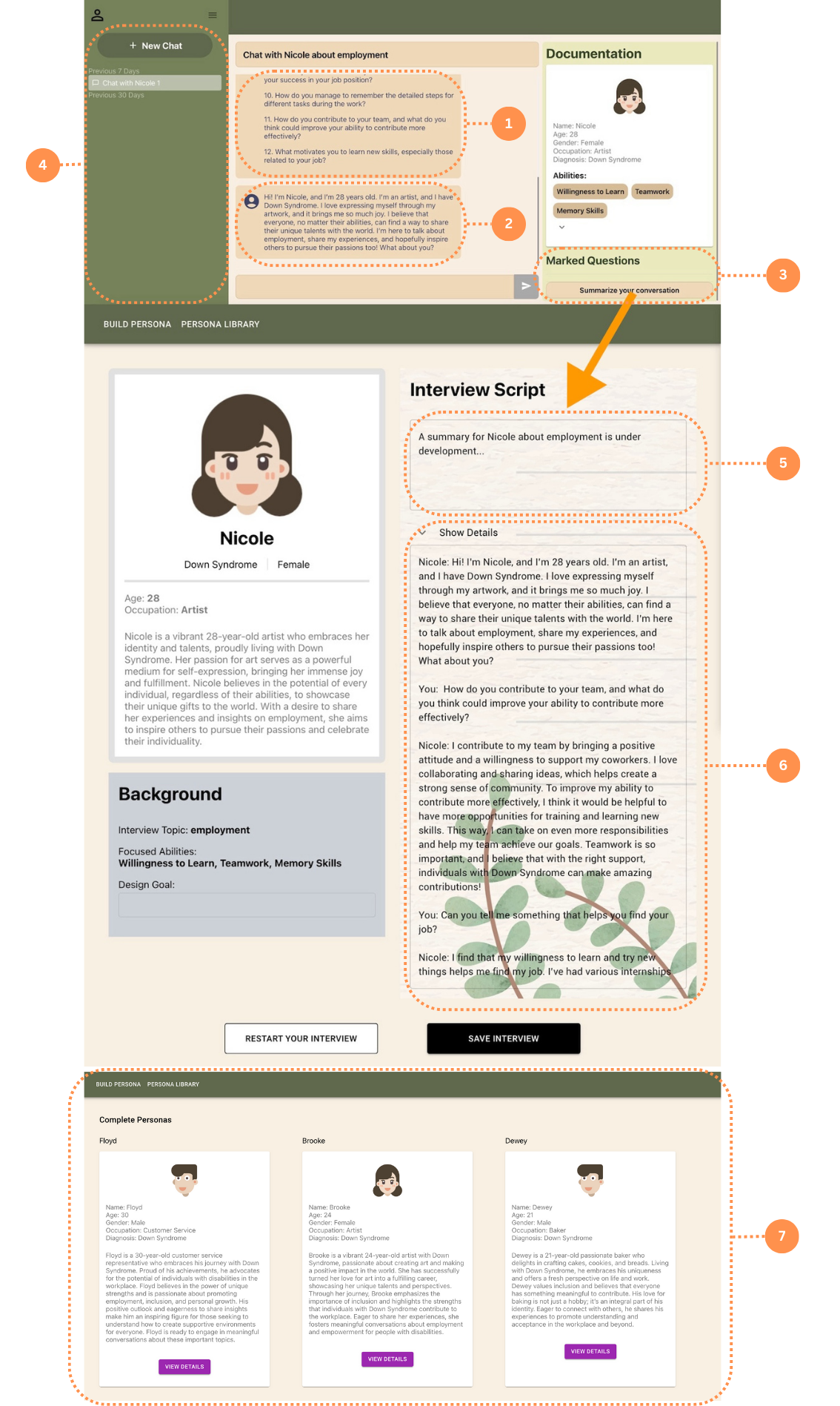}
    \caption{Interaction Phase - Features: (1) Pre-defined questions (2) Chat with Persona-L (3) Mark questions (4) Historical chats (5) Summarize questions (6) Detailed Interview scripts (7) Persona Library}
    \label{fig:interaction phase}
\end{figure}

\textit{Vignette:} Jane has a range of Persona-L features at her disposal to facilitate meaningful engagements with created personas, enhancing her research and design process related to Down Syndrome employment strategies.\\ Persona-L supports this objective  with the following features (see Figure \ref{fig:interaction phase} and Appendix \ref{phase3 user stories} for additional details):
\begin{itemize}
     \item \textbf{Pre-defined Questions}: This feature allows users to utilize a set of sample questions that are relevant to the theme/abilities they are focusing on, such as Memory Skills and Teamwork, providing a structured starting point for exploration. 
     \item \textbf{Chat with Persona-L}: This feature enables users to engage with Persona-L by asking open questions. Users can highlight significant questions and filter out less relevant ones, helping them manage important information more efficiently. 
    \item \textbf{Documentation and Annotation:} All questions and changes are logged in an Interactive Timeline. Users can add notes or comments to each logged event, documenting the rationale behind decisions and changes. The advice for a transparent document of the creator's process and motivation can serve as an invaluable resource for refining and improving processes over time\cite{subramonyam2024bridging}. 
    \item \textbf{Historical Chats}: Users have access to previous chat records while being able to initiate new conversations, ensuring continuity in their research and the ability to compare different interaction outcomes over time.
    \item \textbf{Detailed Interview Scripts}: With access to comprehensive details of their interview scripts, users can thoroughly prepare for interactions in person, ensuring a well-rounded collection of data provides a deeper understanding of their target user.
    \item \textbf{Persona Library}: Users can save and manage multiple personas, allowing them to revisit and refine these profiles as their project evolves.
 \end{itemize}

\subsection{Grounding Persona-L in User Data via RAG and Prompting} 
\label{lab:RAG}

\textbf{User Data}: The persona profiles, the theme-specific abilities, and the Persona-L chat responses are grounded in the responses of an LLM trained on collected user data. The data collection process was designed to capture a broad range of experiences from individuals with Down syndrome, using sources like YouTube, the Canadian Down Syndrome Society \cite{CDSS}, World Day Down syndrome \cite{WDSD}, and other advocacy platforms. This user-generated content provided qualitative insights that were then categorized into three key themes: employment, education, and family, to ensure comprehensive coverage of the challenges and abilities of the target group (see Appendix \ref{Methodology}).\\
\noindent
\textbf{Grounding Responses through RAG}: One of the challenges in utilizing LLMs is managing limited datasets and avoiding hallucinations that is caused by the LLM being presented with complex problems that it has little context and direct knowledge of, or when it fails to effectively recall from its extensive training knowledge \cite{huang2023surveyhallucinationlargelanguage}. To address this, we employed a Retrieval-Augmented Generation (RAG) approach. RAG enables the LLM to retrieve relevant, curated data in real time from a grounded dataset before generating a response. This ensures that the system leverages the latest and most accurate information when crafting persona descriptions and responses, significantly reducing the potential for hallucinations or inaccurate content. We leverage In-context Retrieval-Augmented Language Model(RALM) as one of the RAG methods proven to be efficient in mitigating hallucinations of the LLMs by providing chosen related information in the prompts of the LLMs \cite{ram-etal-2023-context}. By connecting the LLM (GPT-4o mini) to a validated dataset, RAG helps ensure that all dynamic interactions in Persona-L are contextually accurate and grounded in real-world data.\\ 
\noindent
\textbf{Prompting Strategies for Persona Generation and Interaction/Conversation Phase}
GPT-4o mini was chosen as the most viable LLM for Persona-L because it is well-suited for generating rich, context-sensitive content without extensive retraining \cite{brown2020languagemodelsfewshotlearners}. For persona generation, our prompts follow proven prompting principles inspired by \cite{schuller2024generating}. The general guidelines include providing context, specifying requirements, and using delimiters to clearly define different parts of the prompt. After we receive the required qualities of the personas, the scenarios involved, and the inputs by the users from the front-end (such as selected demographics, questions in Interaction/Conversation Phase), we first filter related extracts of documents and knowledge from our database with the RAG before we generate content. Then we employ the aforementioned In-Context RALM method to generate prompts containing related information and persona qualities and pass on the prompt to the LLM to generate responses that fit with the persona profiles. For example prompts for LLM persona generation see Table \ref{tab:example_prompts} in Appendix \ref{LLM persona generation prompt example}.
\section{User Study} 

\subsection{Participants}
We recruited six participants with backgrounds in UX and HCI who are familiar with the use of personas in UX design (see Table~\ref{tab:participants} for demographic information). The recruitment process included a screening stage where participants completed a questionnaire about their occupation, years of experience in UX or HCI research, experience with creating and using personas in their design/research practice, and familiarity with ChatGPT. This screening ensured that participants had a solid understanding of personas and their application in the UX and HCI fields.

All participants had experience with creating or using personas in UX design or research and had also used ChatGPT, either occasionally or regularly. Their familiarity with personas was crucial for providing informed and relevant feedback on our system. The participant group was selected to represent a range of experiences and roles within UX and HCI fields to provide diverse perspectives in the evaluation of the user interface. The study was approved by Institutional Review Board (IRB).

\begin{table}[H]
\caption{Participant Demographics and Experience Summary} \label{tab:participants}
\centering
\begin{tabular}{|l|l|l|l|l|l|}
\hline
\textbf{Participant ID} & \textbf{Age} & \textbf{Gender} & \textbf{Country} & \textbf{Occupation} & \textbf{Years of Experience} \\ \hline
P1 & 25-34 & Female & Canada & Product Designer & 3-5 \\ \hline
P2 & 35-44 & Male & USA & UX Designer & 1-3 \\ \hline
P3 & 25-34 & Female & USA & UX Designer/UX Researcher & 1-3 \\ \hline
P4 & 25-34 & Male & Canada & HCI Lecturer & 3-5 \\ \hline
P5 & 25-34 & Male & USA & UX Designer & >5 \\ \hline
P6 & 35-44 & Female & USA & Senior Product Manager & >5 \\ \hline
\end{tabular}
\end{table}

\subsection{Method and Data Collection}
We conducted six individual interviews with participants using Zoom. At the beginning of each interview, participants were provided with a brief overview of the project. Participants were asked about their previous experience with personas. Then, we presented them with five design scenarios and asked them to choose one of these scenarios to build a persona using  Persona-L  (see Table~\ref{tab:scenarios} for details). Participants were instructed to identify at least two pain points and two design opportunities. As they completed the task,  participants were asked open-ended questions (see Table~\ref{tab:interview_questions} for details). 

All interviews were recorded and transcribed. Thematic analysis was utilized to analyze the qualitative interview data, following the guidelines provided by Braun and Clarke \cite{braun2012thematic}. The process involved familiarizing ourselves with the data through multiple readings of the transcripts, coding the data using NVivo 12, and developing themes. Two coders (paper authors) coded 25\% of transcripts and agreed on the code book, after which one coder completed coding the rest of the transcripts. At each stage, both coders discussed the emerging themes and agreed on the code book.

\begin{table}
    \centering
    \hyphenpenalty=10000
    \exhyphenpenalty=10000
    \begin{tabular}{|l|p{0.14\textwidth}|p{0.36\textwidth}|p{0.28\textwidth}|}\hline
         Theme&  Situation&  Design Task& Design Goal\\ \hline 
         Employment&  Seeking for a Job &  You are interested in understanding what can be helpful for individuals with Down syndrome when seeking employment. Use the persona to explore their experiences in job searching, the support they receive from individuals and social groups, and their recommendations.& The goal is to gain insights into effective communication strategies and support systems that can assist individuals with Down syndrome in their job search and workplace integration.\\ \hline 
         Employment&  Career Development &  You are interested in learning about career advancement opportunities for individuals with Down syndrome. Use the persona to explore their experiences with various training programs, mentoring relationships, and career advancement strategies that have supported their professional growth.& The goal is to propose products or services that facilitate effective communication and mentorship opportunities to support career development and long-term engagement for employees with Down syndrome.\\ \hline 
         Employment&  Enhancing Work Efficiency &  You are interested in exploring how different products can enhance work efficiency specifically for individuals with Down syndrome. Engage with the persona to understand the tools, technologies, and products that have significantly improved their productivity and job performance.& The goal is to gather information about potential communication features or products that could be developed or adapted to help individuals with Down syndrome work more effectively and independently.\\ \hline 
         Education &  Learning at School&  You are interested in understanding the learning experience in a school setting for individuals with Down syndrome. Engage with the persona to explore the most effective learning methods, challenges faced, and strategies to overcome these challenges.& The goal is to design educational tools or curricula that are better suited to enhancing communication and learning experiences for individuals with Down syndrome.\\ \hline 
         Family&  Interacting with Family&  You are interested in understanding the day-to-day interactions within the family setting for individuals with Down syndrome, particularly how they communicate with family members. Leverage the persona tool to explore communication methods and the use of technology in maintaining family relationships.& The goal is to design more inclusive and supportive family communication tools or strategies.\\ \hline
    \end{tabular}
    \caption{Interaction Scenarios for user study. This table outlines specific scenarios under the themes of employment, education, and family, detailing the situations, interactions, and goals of engaging with personas representing individuals with Down syndrome.}
    \label{tab:scenarios}
\end{table}

\begin{table}[ht]
    \centering
    \begin{tabular}{|c|p{0.6\textwidth}|p{0.2\textwidth}|}\hline
        Number & Evaluation Statement & Evaluation Metric \\ \hline
        Q1 & Tell us a little bit more about your experience with personas, how do you create them, and how do you use them? What is involved in each process? & Background \\ \hline
        Q2 & Have you ever used personas that were built by others? & Background \\ \hline
        Q3 & How familiar are you with designing for people with Down Syndrome? & Background \\ \hline
        Q4 & Where do you struggle the most when designing for user groups you are less familiar with? What would help you in that process? & Background \\ \hline
        Q5 & Compared to traditional methods of persona creation, what features of our interface do you find most practical? Can you provide an example that illustrates the utility of these features in enhancing your interaction with the personas? & Trust \\ \hline
        Q6 & Can you tell us a bit more about your process and your experience using this tool? & Overall Experience \\ \hline
        Q7 & How do you choose a persona based on its description? & Persona Description \\ \hline
        Q8 & How do present abilities impact your design process? How do they impact your understanding of people with Down Syndrome? & Abilities \\ \hline
        Q9 & How did suggested questions, if at all, contribute to your process? & Suggested Questions \\ \hline
        Q10 & Can you describe how interacting with the persona has influenced your design process for these personas? Please provide an example of a design decision that was directly influenced by feedback from an interactive persona. & Interaction \\ \hline
        Q11 & Compared to predefined Persona Profiles, do you think the interaction process with interactive personas allows you to better understand individuals with Down Syndrome? Why? Please provide examples to illustrate this. & Interaction \\ \hline
        Q12 & Can you describe a scenario where an interactive persona provided insights that you believe a predefined Persona Profiles would not have revealed? & Interaction \\ \hline
        Q13 & How informative are the responses you are getting from the persona? & Quality of Responses \\ \hline
        Q14 & Did you get the impression that the persona consistently exhibits its defined character traits and maintains its backstory throughout all interactions? & Consistency \\ \hline
        Q15 & How helpful would this tool be in your design practice in the future? At what stage in your design process would this be most useful (persona creation, ideation, decision making, communication)? & Perceived Helpfulness \\ \hline
        Q16 & Would you trust this persona in your design process? & Trust \\ \hline
        Q17 & Any other thoughts you'd like to share with us? & Overall Feedback \\ \hline
    \end{tabular}
    \caption{Interview Questions}
    \label{tab:interview_questions}
\end{table}

\section{Findings}

\subsection{Challenges in Working with Personas from Prior Experience} 
Participants shared their previous experiences and the challenges they encountered when working with personas, particularly when working on personas for users they felt less familiar with or personas built by others. Several themes emerged as significant challenges in persona creation and utilization, including lack of access and time constraints, insufficient information and context, difficulties in building empathy, and concerns about trust and credibility.

\subsubsection{Lack of Access to Users and Time Constraints}
Several participants highlighted the difficulties in creating accurate personas due to limited access to targeted user groups and time constraints, particularly in the context of school projects. In these situations, participants admitted that they sometimes fabricated personas based on assumptions rather than thorough research or interviews. 
A few of the participants also noted while working with people with complex needs, access to user data or empirical data is restricted due to practical and ethical reasons. For example, P4 noted the challenges in working with users who have disabilities, particularly Down syndrome: "Sometimes there's also, I suspect working with people with disabilities, especially Down syndrome, it would not be [a] normal user session. There have to be a lot more accommodations needed, and this [Persona-L] I think moves that and makes it a bit easier for us to access their knowledge to make applications for them." This highlighted the importance of access to knowledge from user groups, which can be difficult to obtain in conventional user sessions.
In addition, as P1 stated, "I guess with real humans, it would be a little bit more time-consuming... because you have to schedule appointments with them and you have to let them know would you be available any time of this project." In fast-paced software development processes, designers usually have access to user interviews only at the beginning of the project while creating personas, and it is really hard to revisit and ask additional questions to the user in the design phase. P6 noted, "I totally understand where there could be utility; it could be beneficial, like a positive impact on the research and tech community if it helps get a perspective into vulnerable groups or hard-to-access groups that generally get disregarded in the design and product decision process, like Down syndrome, neuro-divergent, et cetera. I think that could have a positive impact. It’s like making it easy for me to get some input rather than just saying, 'I don’t have the time to make my product more inclusive.' So I think there’s a positive societal impact potential." 

\subsubsection{Lack of Information and Context when Working with Existing Personas}
Participants also expressed concerns about the relevance and applicability of the information included in personas. P3 shared that they often found it challenging to determine how to use the data provided in personas, especially when the information felt overwhelming or disconnected from the design tasks. P3 noted, "Sometimes it's overwhelming when I get handed personas that have a lot of information... I end up not using the persona." This highlights the need for more contextual relevant and streamlined information included in personas. 

Participants expressed it has been hard to build empathy with personas that are created by others, especially if they are less familiar with the user groups. Participants noted that it was difficult to connect with and fully understand the needs of user groups they were less familiar with. P4 mentioned, "If the developers are not familiar with the demographic, they typically lack understanding of what their needs are. For example, if someone is visually impaired, it's very hard for us to imagine a world where we can't see." Similarly, P6 also expressed "When you use personas developed by others, because I'm not the one who has conducted the primary research, the personas don't feel quite as real to me." 

\subsubsection{Trust and Credibility Concerns}
Another permeating challenge identified by participants was the trustworthiness of personas, particularly when working with personas built by others. P6 expressed concerns about the credibility of personas developed by others, particularly when the primary research was not done by themselves "There's a little bit of a question in my mind of how much should I trust these personas? How rigorously did they conduct the research before they developed these personas?" This raised a trust issue when users are not directly involved in the data collection process or persona creation process. 

\subsection{Previous Use of LLMs in UX Design and Research sets the Context for Persona-L}
All participants were familiar with LLMs, particularly ChatGPT, as they regularly use these tools in their work when dealing with large datasets or vast amounts of information that need to be processed. For example, P2 shared their experience using ChatGPT for data analysis during tech conferences, where they interviewed around 20 people, resulting in a significant amount of information to process. They found that inputting interview data into ChatGPT allowed them to quickly generate summaries and categorize the information into themes, which helps them manage and understand the large dataset efficiently. Similarly, P6 also shared an example of how ChatGPT is currently being used in research. They described a process where an AI tool, such as Outset, is used for recruiting participants and moderating interviews, with transcripts subsequently being summarized in ChatGPT to identify trends.

This prior experience with LLMs, especially ChatGPT, shaped participants' expectations and understanding when interacting with Persona-L. Participants frequently drew comparisons between Persona-L and ChatGPT, noting similarities in interface and functionality. P4 shared "I think the resemblance of this [Persona-L] to chatGPT is one of the reasons perhaps that I think there was a previous existing knowledge in my mind of how to interact with it." P3 and P4 both mentioned that familiarity with ChatGPT's interface contributed to their comfort level in using Persona-L. 

\subsection{Persona-L Easy Access - "You can ask questions anytime, anywhere"}
A key theme that emerged from the interviews was the accessibility of Persona-L. Participants emphasized how the ability to interact with personas at any time and from any location offered a significant advantage over traditional methods. As P1 articulated, "It is quite accessible. You can ask this persona anytime, anything, so you can ask them in ideation and you can ask them in the last phase, which is testing. It is very accessible." Our findings indicate that in fast-paced industries, where re-interviewing users or revisiting earlier phases is often impractical, there is a need for enabling designers to gather insights and revise personas at various design stages as the design goals shift. In support of that, P1 elaborated on the use of Persona-L: "Since you have access to asking a lot of questions,... Given how accessible it is, you can always like, okay, I remember I want to ask this question now so I'm going to come back and ask it. But in a more realistic setting, that part is always kind of hard to go back to. You create personas and you're moving on to the next phase and realistically, a lot of workplaces and companies, they're not going to go back and be like, okay, let's change this up. Most of the time they just want to move forward because there are deadlines. But with this, I think it's more accessible overall."

Participants highlighted how the persona made it easier to engage with user groups that typically require more accommodations, such as individuals with Down syndrome. One participant highlighted Persona-L's potential to improve accessibility in the design process, noting, "Interacting with someone who has Down syndrome... streamlines the communication... making it easier for us to access their knowledge to make applications for them." (P4)

Some participants also found Persona-L to potentially foster better collaboration within design and development teams due to its accessibility. P3 mentioned that team members, such as product managers and engineers, who may not have direct interaction with end-users, could still gain valuable insights and develop empathy for the target user group by having a five-minute conversation with Persona-L. This shared understanding can avoid miscommunication and ensure that all members are aligned with the user-centered goals of the project. P2 also suggested that Persona-L could enhance on-boarding for new team members due to easy accessibility, allowing them to quickly familiarize themselves with personas through interactive chats. P2 explained, "I just let them use this tool to chat with him [Persona-L] to learn about this persona, instead of reading about different personas...."

\subsection{Enhancing Persona-L Consistency and Context through Ability Drivers and Blockers}
Often personas are seen as lacking context. Our findings indicate that providing ability drivers and blockers, which are consistently reinforced throughout interactions with Persona-L, helps maintain a continuous understanding of the persona's context. This understanding of context was found useful by participants potentially during the ideation and requirement gathering stage, as it provided them with a deeper understanding of user needs and challenges as discussed below. 

\subsubsection{Ability drivers and blockers deepen understanding of users but need to be balanced with pain points}
Detailed descriptions of ability drivers and blockers were seen as valuable in providing a deeper understanding of the persona's capabilities. P3 noted that "I like the ability blockers a lot... help me understand their challenges", while P2 found that these abilities give "more detail about the user's capability. It’s helpful." P1 also shared "The one thing that stuck out to me was these stories that you can hover on... it did make me feel a little bit overwhelmed, but it was also helpful for context." These insights suggest that including specific ability drivers and blockers with stories can provide more comprehensive insights into the abilities and disabilities of people with Down Syndrome.

While participants find the abilities section helpful, they also raised some concerns about the focus on positive traits, which could lead to an incomplete understanding of the users. P4 suggested that incorporating negative traits, such as being "always late with their submission" or "bad time management," would create a more balanced and realistic view of the persona. P4 emphasized, "We are human, we're not perfect... I think it definitely adds more depth to the personality..." and "having a full circle of personality traits" is crucial for making persona contextual rich, relatable, and effective. P4 also discussed the importance of understanding the diverse backgrounds and histories of personas to avoid one-dimensional representations. P4 noted "Understanding the history or the context of our personas, giving a bit more personality rather than just a very one-dimensional character, giving some depth of different attributes sometimes adds a bit more understanding to them as well." 

Additionally, P6 shared related concerns by discussing the importance of identifying pain points in personas. They explained, "We always talk about in-product, ... Is your product a painkiller or is your product a vitamin? If they don't have pain, then that doesn't inspire me to go find the right painkiller to address their pain. But that is actually very problematic to me. I would rather have Phil be someone who complains a lot because the users and the personas that can succinctly summarize all of the complaints help guide me to what the solution could look like." This perspective highlighted the importance of capturing both positive and negative traits in Persona-L to inspire meaningful solutions and make the personas more useful in the design process. 

\subsubsection{Consistency in Responses ensures Participants they are talking to the same Persona}
Participants found Persona-L's description and responses to be consistent. P1 highlighted that the consistency between Persona-L's predefined abilities and their responses during the conversation helped maintain the perception that they were interacting with the same Persona-L throughout. They shared, "When this person is answering, you already have a kind of background of these skills and abilities. So maybe that is also helping with the consistency bit where I feel like these are aligned with what I just read, so these generated responses are aligning and they are consistent."

Similarly, P4 observed that the persona repeated key concepts during the conversation, which reinforced Persona-L's knowledge base and made the interaction feel consistent and trustworthy. They noted, "It seemed that, I think at some point it talked about visualizing in one of its answers and it repeated that, reiterated that down here. So I think it is consistent, it has the knowledge base.” P5 also enjoyed the conversational flow and stated that "the flow of conversation is actually very impressive. It's actually able to refer to what we've talked earlier before." This consistency in Persona-L’s descriptions, abilities, and dialogue allowed participants to experience a continuous and coherent interaction, which is essential for providing effective context and building trust.
While participants found Persona-L to be consistent, some participants also expressed concerns about repetitive content in Persona-L's responses. P1 mentioned, "It does feel like this is a bit more repeating than the actual interviews because I think interviewees do also, not always, but they do to a certain extent are aware of what they're talking about where they try not to repeat." This suggested while consistency in contents and responses is valuable, it should also be balanced to avoid redundancy, which could create friction in user experience.

\subsubsection{Potential for Leveraging Persona-L in Ideating and Requirements Gathering}
Several participants found  Persona-L useful in the requirement gathering stage. P4 mentioned that they would use this tool "in requirements gathering trying to understand the needs and issues and struggles that a person with downstream would have." P1 also shared "When you're ideating, you are just kind of throwing ideas out there to see what sticks. I feel like with this generative content, if I have an idea, I would kind of try to ask questions about it and get more context, see if it would work." Persona-L with its generated content would help users to get more context and therefore assist in the design process and allow designers to make more informed decisions. 

However, several participants suggested that the content could be more digestible if presented in a structured format, such as bullet points. P5 specifically asked, “Can it be in bullet points?” Similarly, P1 found the generated content lengthy and difficult to digest. They shared, “I wish there’s a bit of maybe structure in the content… bullet points or highlights.” Although the conversation summary feature was initially considered for to provide a more structured and easily digestible summary of the conversation, this feature was still under development at the time of the interviews.

\subsection{Persona-L Supporting User Connection and Empathy - "She's asking me a question?"}
Our findings show that using Persona-L in conversational interactions enhanced participants' empathy and connection. The dynamic exchange fostered a sense of engaging with a real person rather than a system. By asking questions, Persona-L transformed the interaction from an information-gathering process to a more human-like conversation, making the persona more memorable and the experience more meaningful.

\subsubsection{Increasing the depth of Understanding through Chat Interaction}
Participants noted that engaging in a conversation with Persona-L, rather than merely reading persona description text, enhanced their understanding and sense of connection. P2 mentioned that having a conversation with the persona made the information more digestible: "I think it's easy for me to process, to absorb this information, not just give me a bunch of things to read. Here's just a conversation. I gradually absorb this information and I feel that is a real person chatting with me." P2 also shared "By talking to the people (Persona-L), like a real people, I will remember them." This interaction made Persona-L more memorable to participants and increased their sense of connection and empathy. 

\subsubsection{Persona-L asking Question shifts the Dynamic from an Interview to a friendly Conversation}
The fact that Persona-L asks questions during the conversation was highlighted by several participants as a factor that made the interaction feel more human. For example P2 shared, "It's interesting because she's asking me a question. Normally, I just get answers from them, right? Like ChatGPT... but it makes me feel it's a person talking to me, not just answering my question." This exchange of dialogue, rather than one-way communication, makes the Persona-L feel more realistic. Another participant, P4, also found it interesting that Persona-L "always asks me what I think," which made the conversation more natural and interactive. In contrast, because the persona is asking the user questions, the interaction feels less like an interview and more like a conversation. P4 shared  tensions when asked questions:  "It does make it a bit more of a natural conversation... but at the same time, I'm here to extract information, not provide information." This finding indicates a need to balance between creating a natural interaction that fosters empathy and ensuring the conversation is focused on gathering insights. 

\subsubsection{Humane and Friendly Tone of Persona-L Promotes Natural Interaction and Empathy building}
The tone of Persona-L was perceived as humane and friendly, which further contributed to building empathy and user connection. P1 shared "I think the tone is very friendly. The tone is quite humane," indicating that the persona's communication style was effective in creating a positive and relatable user experience. Additionally, the natural flow of conversation where Persona-L's response is situated in the context, was appreciated by the participant. P5 shared "The flow of conversation is actually very impressive. It's actually able to refer to what we've talked earlier before."(P5) Even though some responses were more detailed than one might expect in real-life conversations, the overall effect was that of a natural and engaging interaction. One participant noted, "It seems like a natural conversation... I said maybe a word and it responded in sentences... but it does make it natural, which I think is if I were someone who did not interact with ChatGPT and these AI models and just interacted with Devin [persona's name] as if it was Devin, it can definitely be sold to me that this is an individual that is talking to me."(P4)

In addition, P6 critiqued the use of a generic icon to represent someone with Down syndrome, pointing out that it lacked authenticity due to the absence of physiological traits associated with the condition. They stated, “Down syndrome is a condition that has physiological traits. So I don’t think you should use a very generic kind of icon to represent someone with Down Syndrome.” 

\subsubsection{Need for Balancing Response Length and Language Appropriateness to avoid Disconnection}
While participants find the chat interaction helps with empathy building, some participants also noted that the length of Persona-L's responses and the level of language used sometimes created a disconnect. For instance, P4 mentioned that "Devin is using very strong words for a 15-year-old, big words," indicating a mismatch between the Persona-L's age and the language used. Another participant, P1, observed that the content was "wordy and lengthy," which made it harder to digest. P4 also pointed out "I do like the length in terms of the question I'm asking, the answers are being provided to me at a depth that I think is not too shallow but not too deep enough that for me I can understand it and I know I can poke for more information if I need it." However, the lengthy nature of these responses could "makes it a little bit less human" (P4) if not balanced correctly.

\subsubsection{Design tension: Shifting Perspectives}
One participant, P1, noticed that there is tension when shifting from empathizing with Persona-L during the ability phase and transitioning to interviewing Persona-L from the interviewer's perspective in the Interaction phase. P1 shared, "When I was selecting the skills and abilities, I did imagine myself in their shoes ... then when the chatting screen loaded, it almost felt like my brain has to do a flip into, okay, now I'm talking to this person that I was just imagining myself as." Similarly, P2 mentioned that having the Persona-L profile icon in the conversation dialogue to match the Persona-L profile image would provide some clarity. This suggested there may be a need for a mechanism to assist users in transitioning from empathizing with Persona-L by "imaging themselves in the user's shoes"  to interacting with it. 

\subsection{Trusting Persona-L}
Our findings suggest that while users find the information and responses from Persona-L believable, they emphasize that transparency in data sources and subject matter expert (SME) verification are key factors that influence their trust in the tool. Ensuring that users can trace the data back to credible sources, or allow them to input primary data themselves, along with knowing that the data and Persona-L have been validated by SMEs, can help mitigate trust issues and enhance the tool’s reliability. Additionally, regular data updates and transparency about the processes used in Persona-L are crucial for maintaining trust.

\subsubsection{Data Transparency helps build Trust in Persona-L}
Participants consistently expressed a strong preference for knowing the origins of the data that was used to build the persona. Knowing the source of data allows them to evaluate the reliability and accuracy of the information. For example, P2 shared, "I would like to know the source so I can trust it. Especially if the data from the research I conducted... I know that's accurate or not." Another participant (P3) stated, "it'd be nice to see where that data is coming from or at least, maybe it's from a research study that I can maybe look into a paper so that it adds to more credibility." P3 also noted "If I were the person who actually put the data into the database, then I think I would trust it a hundred percent. Even though I know that I could get inaccurate data, I would trust it more if it's coming from a database that I created."
Another participant (P5) expressed that while Persona-L seems believable, it would be beneficial to have references regarding the data. P5 pointed out the value of references in building trust: "AI is not 100 percent correct, but with the response it's giving me, it shows that it's well-trained ... Can this also have references once in a while?" Knowing the data source and having the ability to delve deeper into the data helps users build trust in Persona-L and the information it provides.
P6 further emphasized the importance of data transparency and the need for regularly updating data to provide relevant knowledge: "I would probably have trust issues if I don't feel like there's enough data. And I would also have trust issues if I feel like the data is not being updated... what if the job market has shifted for them? What if there's been new tools or new technologies out, whatever that might be." P6 suggested, "Can you be more transparent upfront... a hundred individuals' data were used to create this LLM and we updated 20 individuals every other week or whatever that is." Keeping data up to date helps prevent changes over time from rendering personas outdated, less accurate, or irrelevant.

\subsubsection{Subject Matter Expert Evaluation and Verification}
Participants also emphasized the importance of human and subject matter expert (SME) evaluation and verification. For instance, P4 suggested the implementation of a "badge" or similar marker to indicate that the specific Persona-L has been verified by a human expert, similar to the verification check marks used for celebrities on social media. P4 shared "I think the idea of the little badge or something that, right beside names that the answers provided by Devin are pulling data from a human-verified dataset or just like you have the verified check mark besides someone who's a celebrity." 
P4 also discussed the importance of trusting both the system and the individual persona. P4 shared, "I would be more trusting of a persona that has been tested by other people because, in my mind, that has been refined in some sense." They proposed that a confidence score or verification badge could be added to indicate how reliable a persona's responses are, which could help users trust the information provided by the persona.

\subsection{Diversifying Persona-L}
Our findings highlight the need for diversity in both persona creation and content or responses provided by personas. Participants emphasized the importance of generating personas that reflect a broad range of demographics, characteristics, and backgrounds to ensure they can represent the diverse needs and experiences of users effectively. 

\subsubsection{Needs to further diversify Persona-L}
During the interview, P2 emphasized the importance of having different representations among personas to increase diversity. When asked about how he/she selected a persona to work with, P2 mentioned the value of diversifying across attributes such as age, gender, and experience levels. This approach helped to broaden the understanding of user groups and needs, as P2 shared "I just try to diversify the generated personas... one male, one female, different ages," to ensure that the personas selected offer diverse perspectives and insights.

In addition, P6 raised concerns about the lack of diversity in the Persona-L-generated personas, noting that the personas felt too similar in their description and lack of distinct characteristics needed to reflect different user experiences. P6 emphasized the importance of having personas that are not only demographically distinct but also differ in their relationship to the product being designed. For example, P6 stated, "One could be Down syndrome and somebody who is really physically active and has no mobility issues and really wants to find jobs that keep them active. But then you could have someone who has Down Syndrome but has mobility issues and therefore needs to have additional jobs that are maybe indoors and are a little bit easier office stuff. That may be a third one someone with Down syndrome who has speech impairment or speech therapy type of issues, and they need a job that maybe they can type and communicate through words, but not through verbally. So that's the distinct aspects of the condition that changes the relationship and changes the relationship to the feature and makes them a slightly different user in that way." 

\subsubsection{Diversifying Insights through Simulated Group Persona Interaction}
In addition to the need for diversifying personas, participants also expressed interest in using Persona-L to engage with multiple personas simultaneously to increase the diversity in insights. P2 mentioned the potential efficiency of having group chats with two or three personas at once, rather than starting separate conversations with each one. This feature, according to the P2 could further streamline the design process and enhance the richness of insights gathered. P2 also mentioned the efficiency of potentially conducting group chats with multiple personas, stating, "I want to know more personas because I don't want to start nine conversations... If I can chat with two or three, then just three conversations to know all the personas." This ability to quickly gather diverse perspectives without the logistical constraints of coordinating with real people was seen as a significant advantage.

\section{Discussion}
\label{lab:discussion}

This study investigates how Large Language Models (LLMs) can be integrated with an ability-based framework to create dynamic and context-rich personas of people with complex needs. Our findings confirm that Persona-L facilitates natural and empathic conversations, which aligns with prior research on LLM-generated personas being believable and relatable \cite{salminen_deus_2024, zhang_auto-generated_2024}. Additionally, the incorporation of ability drivers and blockers into the persona design provides participants with a more comprehensive understanding of persona capabilities, which mirrors the goals of Ability-Based Design (ABD) as highlighted by Nolte et al. \cite{nolte2022creating}.

\subsubsection{Implementing LLMs in Persona-L} 

LLMs have shown potential in generating believable, relatable, and informative personas. However, inherent biases, stereotypes, and hallucinations persist in LLM-generated personas \cite{schuller2024generating} and LLM-generated data. To address these issues, we employed the In-Context Retrieval-Augmented Generation (RAG) method (see Section \ref{lab:RAG}). This approach helps mitigate hallucinations by grounding the generated content in a human-curated dataset. The RAG system retrieves relevant data from this grounded dataset to ensure that the outputs from the LLMs are both contextually accurate and reflective of real user behaviors and experiences from the dataset. In addition, previous research by \cite{goel2023preparing} pointed out that the responses of LLMs-generated personas are generic and inconsistent after multiple prompts. Our findings indicate, that Persona-L's responses were consistent over multiple rounds of prompting and we relate that result to the use of RAG in Persona-L rather than direct interfacing with LLMs.

Previous research has found data-driven personas creation to be efficient, cost-effective, scalable, and reflective of user behavior \cite{zhang_data-driven_2016,mcginn_data-driven_2008}. Our findings support these advantages, as participants found the Persona-L creation process to be both efficient and easily accessible. However, participants also raised concerns related to data transparency, emphasizing the need for data verification by human subject experts to ensure the accuracy and authenticity of the personas. 

Additionally, our findings highlighted the need for regularly updating the dataset to keep the personas current. Currently, in our system, this process is done manually, which is time-consuming and not scalable. This raises important considerations regarding the best practices for keeping the datasets up to date and expanding over time. It is worth considering whether incorporating real-time online user data, as suggested by Jung et al. \cite{jung_persona_2017}, could keep data-driven personas up-to-date in this context. Alternatively, our findings suggest that enabling users to input and customize their data entries into the Persona-L system could maintain the relevance and trustworthiness of the personas, and presents an opportunity for consideration of how to support collective efforts on persona creation. 

\subsubsection{Building Empathy through Interaction}

Previous literature discussed personas as being abstract, impersonal, and misleading \cite{matthews2012designers}. This concern aligns with our study's findings, where participants raised concerns about the relevance and applicability of the information included in personas. The use of LLMs to enable users to interact with Persona-L through chat helps to build empathy and understanding. \cite{jansen2022create} also presented how understanding the wider context of personas supports feelings of empathy towards users. This dynamic interaction is more aligned with previous research suggesting that LLMs can offer more natural and personalized interactions that elicit empathy and connection\cite{ha2024clochat}. 

Our findings indicate that engaging in a conversation with Persona-L, rather than passively reading persona descriptions, significantly enhances users’ understanding of personas as it allows them to ask follow-up questions. This is consistent with research that emphasizes the need for multifaceted, dynamic personas in LLM simulations, which allow for deeper engagement and better reflection of real users’ evolving needs and behaviors \cite{cheng2023compost, schmidt2024simulating}. The dialogue-based approach in Persona-L enables more iterative and context-rich interactions, which enhances user connection compared to traditional static personas.

The humane and friendly tone of Persona-L, combined with the consistency in its responses, contributes significantly to building connections with users. Previous research has shown that perceived friendliness and realism of deepfake personas increase trust \cite{kaate2023users}. Persona-L’s ability to ask questions shifts the interaction from a one-sided, generic response model to a more consistent, interactive conversation. With the implementation of the RAG system, participants found that Persona-L’s responses were consistent and that the persona could reference previous conversation topics, adding continuity and coherence to the interactions. This approach addresses some of the issues identified in \cite{goel2023preparing}, where LLM-generated personas were rated lower by UX design and development professionals due to a lack of credibility, empathy, and memorability—issues often stemming from generic responses and inconsistencies across multiple prompts. However, our findings also suggested that further investigation is needed regarding fine-tuning the vocabulary and length of responses to match the persona's demographic characteristics to ensure that the responses are perceived as more relatable and believable. This was also indicated in previous research by \cite{barambones2024chatgpt}.

Additionally, our study reveals that conversational interactions with Persona-L make the persona more memorable. This aligns with previous research suggesting that implementing adaptive algorithms, which tailor persona behaviors based on user inputs, can enable LLM-based personas to evolve over time, fostering empathy through these dynamic adaptations \cite{ha2024clochat}. The real-time responsiveness of Persona-L, grounded in the Ability-Based Design framework, adds another layer of authenticity and depth to the interaction, as it reflects the persona’s abilities and challenges in the context of design text.

\subsubsection{Ability-based framework as a step towards expanding contextual depth}
Our findings suggest that the ability-based framework, particularly through the inclusion of ability drivers and blockers in the Persona-L interface and the consistent reinforcement of abilities throughout Persona-L interactions, significantly enhances users’ ability to build a deep understanding of the persona’s capabilities. This approach aligns with the work of \cite{nolte2022creating}, who advocate for shifting the focus from disabilities to abilities, emphasizing the importance of contextualizing these abilities within specific tasks. Additionally, this approach addresses some of the concerns raised by \cite{edwards2020three}, who argue that traditional methods of creating personas often fail to capture the full contextual depth—including social, cultural, and personal factors—that significantly influence how target users interact with technology. This is also confirmed by the previous research that pointed to the importance of a rich description of the user's context (tasks, and environment) in persona creation \cite{jansen2022create,Holtzblatt_2005}. 

However, our findings also highlight a tension between the helpfulness of focusing wider context and the abilities and the need for UX and product designers to identify pain points efficiently that lead to effective design solutions. While the ability-based framework emphasizes capabilities, it sometimes downplays the negative traits or challenges faced by the persona, resulting in an incomplete or overly optimistic representation. This suggests a need for a more balanced approach that not only celebrates abilities but also acknowledges the difficulties that users may encounter, thereby providing a more comprehensive foundation for design. Finally, the summary feature of Persona-L was under development at the time of the study. We anticipate that this may have contributed to the tension between the depth of information presented and the participants' need to summarize key points identified in the chat. This feature will be included in Persona-L in the future.

\subsubsection{Multiple Factors at Play in Diversifying Personas}
As indicated in related Work, generating more personas results in a better representation of diverse demographic groups \cite{salminen_creating_2022}. \cite{Salminen2022how} also shows that increasing the number of personas positively impacts engagement and demographic diversity. However, as Oudshoorn et al. \cite{oudshoorn2004configuring} pointed out, diversity often decreases during persona creation because a limited number of personas are generated, which leads to under-representation of various user characteristics such as age, gender, and race.

Our findings echoed these concerns, as participants who generated multiple personas with the same age, occupation, medical condition, and theme selections noted that the resulting personas felt very similar. This issue is likely due to the limited user data fed into Persona-L's RAG system, which may have caused the personas to appear overly alike. Salminen et al. similarly highlighted the risk of generating personas that lack diversity or become biased, particularly in large-scale persona generation, where insufficiently comprehensive datasets can lead to limited representation of diverse populations \cite{salminen_deus_2024}.

One potential solution is to integrate larger and more diverse datasets to enrich the variety of generated personas. Future work could also explore generating persona sets using Salminen et al.’s methods \cite{salminen_deus_2024}, which can be paired with diversity evaluation techniques. This would help ensure that the generated personas reflect a broad range of lived experiences and incorporate diverse demographic, cultural, and socio-economic backgrounds. As Salminen et al. \cite{salminen_creating_2022} suggest, generating more personas can significantly improve the diversity and inclusivity of persona representation.

\subsection{Design Implications}
Grounded in our findings and reflections, we propose three high-level principles aimed at rethinking how LLMs can help approach the challenges inherent in personas.
\\
\textbf{Principle 1 - The importance of data transparency and validation in LLMs persona creation} 
To build trust and maintain the relevance of personas generated by LLMs, it is essential to incorporate mechanisms that provide transparency regarding the data sources used to create the personas. Allowing users to customize datasets with their data, along with validation of these datasets by human and subject matter experts(SME), can help enhance credibility and ensure that the personas remain accurate and reflective of real-world experience. This aligns with previous research highlighting the need for data transparency in auto-generated personas to maintain trust and authenticity \cite{zhang_auto-generated_2024}.
\\
In addition to dataset validation, our findings also emphasized the importance of human and SME evaluation and verification. Per \cite{salminen_deus_2024} personas generated using GPT-4 and have revealed bias in LLMs-generated personas, particularly related to age, occupation, and pain points, as well as a strong bias towards personas from the United States \cite{salminen_deus_2024}. Based on the observation of the research, researchers suggested verifying the LLM-generated personas using diversity and bias analysis techniques, as well as subject-matter experts to establish external validity. The prompt design would also be able to address and affect some of the issues in diversity, bias, or quality of the persona \cite{salminen_deus_2024}. 
\\
\textbf{Principle 2 - Matching persona voice, tone, and vocabulary to persona demographics}
In the creation of LLM-generated personas, it is important to design conversations that feel natural, contextually relevant, and believable. The length of responses should be appropriate, reflecting the flow of natural conversation. The language level and vocabulary should also align with the persona's demographic characteristics, particularly their age. Persona-L demonstrated the effectiveness of using humane and friendly tones, along with asking questions, to foster dynamic, interactive dialogue. This approach shifts the experience from one-way communication to a more personalized conversation. However, additional attention should be paid to ensure that the complexity of the language matches the persona's demographic information. The alignment between persona voice, tone, and vocabulary to persona demographics would enhance both empathy building and overall believability of the persona. 
\\
\textbf{Principle 3 - Balance abilities and constraints to present a comprehensive picture of persona complexities}
While the ability-based framework in Persona-L effectively emphasizes users’ abilities, it is equally important to acknowledge the challenges or constraints personas face. A well-rounded persona should provide a balanced view, incorporating both strengths and difficulties to accurately represent the complexity of lived experiences. This nuanced representation helps UX and product designers identify relevant pain points and develop effective solutions. In addition to the persona description, ability drivers and blockers, and conversation elements, presenting information in a structured summary, such as a bullet-point format, further enhances user comprehension by making the persona’s complexities easier to digest.

\section{Limitations and Future Work}
We acknowledge several limitations in our study that also inform our future work, which are outlined below.

\textit{Prolonged use of Persona-L} The participants in our study had limited time to use Persona-L in a short design brief (see Table \ref{tab:scenarios}). This constraint may have hindered participants from fully engaging with the persona through multiple interactions. As future work we are planning a longitudinal study that will allow us to understand how users leverage Persona-L over an extended period in their design practice.

\textit{Detecting Potential Stereotypes in LLM responses} While the use of the RAG method reduces the possibility of hallucinations, it does not eliminate the stereotypes inherent in the training data of LLMs. Although stereotype and bias assessments were not included in this study, future research will incorporate methods like the CoMPosT framework \cite{cheng_compost_2023} to evaluate Persona-L generated content for potential stereotypes. By addressing these limitations, we aim to refine the Persona-L system and provide more comprehensive insights into its long-term effectiveness and ethical implications related to using AI for simulating users \cite{schmidt2024simulating}.

\textit{Co-creation and validation of Persona-L} Another direction that we find of utmost importance to explore is how Persona-L can be utilized in co-creation of personas that will involve users and subject matter experts in persona creation and persona response validation, similarly to what has been previously done with people with diverse needs \cite{neate_co-created_2019}, children \cite{warnestal_co-constructing_2014}, and people with disabilities \cite{heitmeier_persona_2023}. Lived experiences are complex and multi-faceted \cite{prpa2020articulating}, and span beyond the descriptions that is shared by the people publicly via forums we gathered data from. Therefore it is important that we also consider other methods for data collection that will allow for the richness of lived experiences to be captured \cite{prpa2020articulating} and included in Persona-L responses. By involving users and subject matter experts in the co-creation and validation of personas, we aim to enhance the accuracy and trustworthiness of the Persona-L, ensuring they better reflect the complexity of diverse lived experiences.

\textit{Matching Persona-L Traits to User Expectations} Our findings indicate that the tone, language and profile image used by Persona-L should match persona's demographics. The length of the responses should be appropriate for the conversation flow to make the interaction feel more natural. In future work, LLMs need to be fine-tuned to generate responses that match demographic expectations. In addition, participants expressed a need for more structured and digestible content, with some suggesting that bullet points or highlights would improve readability. The conversation summary feature that provides a more structured and easily digestible summary of the conversation was  under development at the time of the study. This feature will be developed in future iterations to provide more concise and structured information.

While in this first iteration of Persona-L we haven't made deliberate decisions about Persona Image and we used generic cartoonish images to represent personas, our findings indicate that images do matter. \cite{Salminen_2022_CanUnhappy} has noted that the facial expression used in persona profiles significantly affects user perceptions. Additionally, \cite{salminen_2023_imaginarypersona} indicates how persona's appearance can impact perceptions of trustworthiness and competence. In future work, we will focus on visual identities of personas to increase the trustworthiness and believability.

\textit{Enriching Persona-L over Time} Additionally, Persona-L's RAG system updates information in real-time during interactions, reflecting user inputs immediately. Next steps involve expending the RAG system so that over time, the system learns from user interactions, refining its responses and improving  accuracy of persona responses. This continuous learning process could potentially help maintain the relevance and contextual appropriateness of the personas. Finally, we aim to expand Persona-L to enable users to contribute their data in persona creation paired with traceability of sources that informed Persona-L responses to enhance users' trust in Persona-L.


\section{Motivation and Authors' Disclosure}

The motivation for our research is two-fold:
First, throughout the years of teaching HCI methods in classroom, we were often faced with the challenges of teaching students how to build personas without directly interfacing with people in the early stages of problem discovery. This is usually the case when designing personas of people with complex needs, and where the goal of building personas is primarily educational. This by no means suggests that the lived experiences of people should not be explored and accounted for in persona creation. We argue that novice practitioners (students), not yet skilled in research methods or specificities of user groups can benefit from additional tools in the early stages of Discovery and Problem Definition, which can and should be validated and expanded on in interaction with people. Additionally, while LLMs have shown great promise in persona creation, we are aware of the current challenges in using LLMs to create personas for people with complex needs. These groups, due to their vulnerability in various contexts, require representations that are both accurate and empathetic. We aim to empower our end users—designers and developers with additional tools to have a more comprehensive understanding of users with complex needs and foster empathetic connections with the target group they aim to serve. Presented research is just a first step toward this larger goal.
By conducting this research we are mindful of the complexities this topic poses and are committed to supporting people with complex needs by creating a tool that can be leveraged to minimize extractive practices often imposed on the subjects. None of the authors identify as people with complex needs, and by no means do we advocate for prioritizing the use of Persona-L in cases where access to people is available, ethical, and benefits the communities in the research foci. The long-term vision for Persona-L as a collaborative, co-creative tool mandates thoughtful evaluation of benefits and impacts to different user groups, which will be addressed comprehensively in future work.
\section{Conclusion}

Persona-L represents a novel approach to creating interactive personas utilizing Large Language Models (LLMs) within an ability-based framework to address the complexities of representing people with complex needs. This approach aims to create more dynamic, context-rich personas that enhance empathy and understanding among UX designers and product developers. Our findings indicate that Persona-L offers significant advantage in enabling users to engage in natural conversation interaction with personas, which deepens a sense of connection and realism. 

However, this study also reveals some challenges, particularly around the need for data transparency and the importance of providing a more balanced presentation of abilities with constraints. These aspects are essential to ensuring that personas accurately represent the complexities of users' lived experiences. Our research provides insights into how designers can use tools like Persona-L to create more context-rich personas that evolve through user interaction. As our research suggests, future iterations of Persona-L will address challenges related to stereotype mitigation and bias monitoring to ensure that the personas remain inclusive and representative. We hope that these insights will guide future efforts in HCI and persona research, contributing to the design of more authentic, empathetic, and context-rich personas.

\bibliographystyle{ACM-Reference-Format}
\bibliography{0.Persona_v01}
\newpage
\appendix
\section{Appendix - Methodology} \label{Methodology}

Our methodology included designing the web-based interface, integrating an ability-based framework, collecting data about the user group to ground LLM responses, training  and integrating LLM model into web-based interface. 

\subsection{Design Tasks}
\begin{itemize}
    \item \textbf{Design and Implement an Interactive LLMs Persona Tool:}  develop a user-friendly interface that enables users to interact with and customize personas using LLMs.
    \item \textbf{Incorporate an Ability-Based Framework:} Our target population for this approach includes individuals with Down syndrome. Traditional persona methods often misrepresent people with complex needs by labeling them through their impairments, the ability-Based Framework can enable to consider the specific abilities and needs of people with complex needs\cite{nolte2022creating}, ensuring that personas can be customized to reflect more realistically their capabilities. 
    \item \textbf{Allow Personas to Simulate Based on Wider Contexts:} Currently, personas are often reduced to a limited amount of information and lack contextual depth \cite{edwards2020three}. We enable users to choose different contexts, thereby creating personas that can adapt and interact within those varied contexts.
    \item \textbf{Integrate LLM Model into HCI Interface:} This integration aims to ensure a seamless user experience by enabling dynamic interaction between the user and the persona creation tool. 

\end{itemize}

\subsection{Procedures} 
\subsubsection{\textbf{\textbf{Data Collection\\}}} 

\textbf{Dataset Gathering: Collecting Stories shared from People with Down Syndrome to ground LLM response generation}
Our data sources include videos and textual content from platforms such as YouTube, National Down Syndrome Society (NDSS), World Down Syndrome Day and similar websites where individuals from Down syndrome communities share their stories. These platforms provide a rich tapestry of personal stories and experiences that are essential for training LLM and grounding LLM responses.

From these stories, we manually extract data (details provided in Table \ref{tab:oneData}), including demographic information such as age, gender, and education level. Additionally, we document their abilities and instances where they outperform average individuals, as well as the challenges they face in various scenarios and contexts. Also, the data collected is categorized into predefined themes significant to individuals with Down syndrome, including education, employment and family. This systematic data collection is instrumental in understanding the specific needs, preferences, and contextual factors of our target groups. The gathered data is utilized to train our LLM model, enhancing its ability to generate more accurate and contextually relevant personas.

\textbf{Dataset Volume}: We have manually collected 80 data records, each representing a real-life story of an individual with Down syndrome. This curated collection aims to encompass a wide array of experiences, highlighting both the unique challenges and accomplishments of each individual.

\textbf{Dataset Nature}: The data is qualitative, comprising textual information that provides in-depth insights into the lives of those featured. This rich textual content includes narratives, personal anecdotes, and detailed descriptions, enabling us to capture the nuanced realities and diverse experiences of individuals with Down syndrome. This approach ensures that our dataset is not only comprehensive but also deeply resonant and contextually relevant for persona development.

\begin{table}
    \centering
    \begin{tabularx}{\textwidth}{|l|X|}
    \hline
         Name& Austin Underwood\\ \hline 
         Diagnosis& Down Syndrome\\ \hline 
         Resources& https://ndss.org/success-stories?page=4\\ \hline 
         Age& 40\\ \hline 
         Gender& Male\\ \hline 
         Region& US\\ \hline 
         Education & Occupational Training Program at Eastern New Mexico University\\ \hline 
 Occupation & President of Austin's Underdawgs  restaurant\\ \hline 
 Family Situation &Supportive parents\\ \hline 
 Label&Independent Living/Employment\\ \hline 
 Abilities &Independent Living: He was thrown into living situation where he had to apply the skills he had learned growing up... 
Employment: He has shown a strong work ethic and adaptability...\\ \hline 
         Challenges & He is unable to do things like read or drive a car... \\ \hline 
    \end{tabularx}
    \caption{Example of One Data Entry Extracted from Real People's Stories }
    \label{tab:oneData}
\end{table}

\textbf{Ground Truth Questions:}
In our methodology, specific ground truth questions are designed to probe deeper into the scenarios derived from the dataset. These questions are crafted to elicit detailed and contextually relevant responses, essential for refining our personas. Based on the real-life experiences of individuals with Down syndrome, we produce answers to these preset questions, ensuring that the responses are both accurate and relevant. These generated responses are subsequently examined and refined by experts to ensure the quality of these answers. Each question is crafted to address key dimensions identified in the scenario summaries, ensuring they are both pertinent and comprehensive. This approach helps in generating responses that are not only informative but also resonate with the actual experiences of individuals with Down syndrome.

\subsubsection{\textbf{\textbf{User Interface Design and Development\\}}} We utilize React.js as our front-end framework, chosen for its dynamic and interactive capabilities. This interface is designed to allow straightforward customization and interaction with personas, ensuring users can effectively shape and utilize them as needed. 
Within this user interface, we provide users with a range of options and offer the following \textbf{\textit{\textit{parameters}}}:\\
a) Demographic Attributes: This includes age, occupation, medical condition etc., which are essential for shaping the persona. \\
b) Theme: Users can choose various themes such as employment, family, and education themes that might influence persona responses. \\
c) Ability: Users can choose a variety of key abilities based on the specific theme.

To be specific, users initially build the basic persona profile by selecting information such as demographic details. After that, the interface will display the key abilities of the persona. Subsequently, users can further interact with the persona by asking follow-up questions. This intuitive, user-friendly interface is designed to provide flexibility in creating personalized personas and support easy and consistent interactions with LLMs-based personas.

\subsubsection{\textbf{Integrating LLM Capabilities with the User Interface\\}}
To enhance the integration of the Large Language Model  with our User interface, we established a following \textbf{Integration Procedures}:
\begin{itemize}
        \item \textbf{Establishing Integration Guidelines.} Both teams collaborate to establish a set of guidelines for integrating the LLM with the HCI interface. This involves defining data exchange formats, API specifications, and interaction flow between the frontend interface and the LLM backend.
        \item \textbf{Iterative Testing and Feedback Loop. }This phase includes setting up the frontend interface to interact effectively with the LLM, ensuring that the user inputs are correctly processed and that the persona responses are dynamically generated. An iterative testing process has been implemented, where the integrated system is tested under various scenarios to identify any issues.
        \item \textbf{Usability Enhancements.} Based on feedback from the tests, further enhancements has been made to improve usability, such as adjusting the interface design, increasing response generation speed, or enhancing the capabilities of the LLM.
        \item \textbf{Final Implementation. } In this phase, the final version of the integrated interface has been launched. Continuous monitoring will be established to ensure the system operates smoothly and efficiently.
    \end{itemize}

 \textbf{API Specifications}
    \begin{itemize}
     \item \textbf{Persona customization phase:} Users input attributes through the \textbf{Persona Creation API} to create a persona profile and generate a description of the persona. To be specific, users input attributes such as theme, name, age, and occupation through the Persona Creation API. This API then processes these inputs to create a detailed persona profile and generate a comprehensive description. For example, when a user inputs the theme "Employment" and specifies the profile with the name "Emily", age "34", and occupation "School assistant", the API returns a persona description that highlights Emily's characteristics and job-related details. \\This tailored response includes a system prompt and an initial assistant message to facilitate immediate interaction, ensuring the persona is ready for engagement based on its defined traits. For a detailed breakdown of the input and output formats of this API interaction, see Table \ref{tab:Persona Creation API}.
        \end{itemize}

\begin{table}
    \centering
    \begin{tabular}{|c|p{0.6\textwidth}|l|} \hline
         Request Body& \{
  "theme": "string", 
  "profile": \{
    "name": "string", 
    "age": "integer",
    "occupation": "string"
    "Medical Condition": "string"\}
\}\\ \hline 
         Request Example& \{
  "theme": "Employment",
  "profile": \{
    "name": "Emily",
    "age": 34,
    "occupation": "School assistant"
    "Medical Condition": "Down Syndrome"
  \}
\}\\ \hline 
         Response Body& \{
  "description": "string",
  "system\_prompt": "string",
  "assistant\_message": "string"
\}\\ \hline 
         Response Example& \{
  "description": "Hi! I am Emily, a 34-year-old school assistant with Down syndrome. Despite my condition, I passionately love my job and is known for my compassionate and diligent nature, consistently fulfilling my responsibilities at the school.",
"system\_prompt": "You are Emily, a school assistant with Down syndrome. Despite your condition, you are passionate about your job and dedicated to your responsibilities.",
  "assistant\_message": "Hello, I'm Emily. How can I assist you today? " \}\\ \hline
    \end{tabular}
    \caption{Persona Creation API}
    \label{tab:Persona Creation API}
\end{table}
\begin{itemize}
 \item \textbf{Persona interaction phase:}  Users input questions and engage in dialogues through the \textbf{Persona Chat API}. The API utilizes the context of the conversation, including previous interactions and the persona's detailed profile, to generate responses that are reflective of the persona's identity. For instance, when a user asks Emily, "What motivates you to learn new skills, especially those related to your job?" the API responds with a message crafted to reflect Emily's motivations and aspirations, emphasizing her dedication and the personal benefits she gains from learning new skills. This interaction is not only tailored to the specific inquiries of the user but also aligns with the persona's background and the scenario provided. The responses are dynamically generated by the LLM, ensuring that each interaction is consistent with Emily's characteristics as a compassionate and diligent school assistant with Down syndrome. This dynamic conversation capability enhances the user experience by making the persona feel more realistic and relatable. For further details on the request and response structure of this API, see Table \ref{tab:Persona Chat API}.

\begin{table}
    \centering
    \begin{tabular}{|c|p{0.6\textwidth}|l|} \hline
         Request Body& {
  "context": [
    \{
      "role": "string",
      "content": "string"
    \}
  ]
}\\ \hline 
         Request Example& {
  "context": [
    \{
      "role": "system",
      "content": "You are Emily, a school assistant with Down syndrome. Despite your condition, you are passionate about your job and dedicated to your responsibilities."
    \},
    \{
      "role": "assistant",
      "content": "Hello, I'm Emily. How can I assist you today?"
    \},
    \{
      "role": "user",
      "content": "What motivates you to learn new skills, especially those related to your job?"
    \}
  ]
}\\ \hline 
   Response Body & \{ "assistant\_message": \{ "role": "assistant", "content": "string" \} \}\\ \hline 
   Response Example& \{ "assistant\_message": \{ "role": "assistant", "content": "I'm motivated by the desire to do my job better and to show my coworkers and friends that I can handle more responsibilities. Also, learning new skills makes me feel proud and capable. It's also a great way to increase my earnings!" \} \}\\ \hline
    \end{tabular}
    \caption{Persona Chat API}
    \label{tab:Persona Chat API}
\end{table}
    \end{itemize}

\section{Appendix: Example Prompts for LLM Persona Generation Strategies} \label{LLM persona generation prompt example}

\begin{table}[h]
    \centering
    \begin{tabular}{|p{0.25\textwidth}|p{0.51\textwidth}|}\hline
         Strategy & Example Prompt  \\ \hline
         General Guidelines & “persona description= ‘At 33, Shea, who has Down syndrome, has shown remarkable courage and determination. . . .’ description Make sure that the personas are consistent and the output is in the form of tables." \\ \hline
         Role-play Prompting & “You are Shea. You are a 33-year-old woman with Down syndrome, bursting with enthusiasm and confidence.”\\ \hline
         One-Shot Prompting & “Target Group: People involved in Down syndrome. Name: Shea. Age: 33. Profession: Retail associate. Hobby + interest: Passionate about disability advocacy, public speaking, and personal growth. . . .”\\ \hline
         Incremental Prompting & “(1) Prompt: Create Personas (2) Prompt: Add the following attributes for each previously created persona: <keywords> (3)
Prompt: ...”\\ \hline
    \end{tabular}
    \caption{Example Prompts for LLM Persona Generation Strategies}
    \label{tab:example_prompts}
\end{table}

\newpage
\section{Appendix: Interaction/Conversation Phase 3 Features and User Stories} \label{phase3 user stories}

\begin{table}[h]
    \centering
    \hyphenpenalty=10000
    \exhyphenpenalty=10000
    \begin{tabular}{|c|p{0.27\textwidth}|p{0.37\textwidth}|}\hline 
         Feature&  Need& Purpose\\ \hline 
         (1) Pre-defined questions&  Jane wants to access sample questions that have been previously asked to LLMs, particularly those related to the abilities she has chosen, such as Memory Skills and Teamwork.& Pre-defined questions will guide users by providing examples of how to frame inquiries based on specific abilities effectively. \\ \hline 
         (2) Chat with Persona-L&  Jane wants to chat with this created persona, to learn more about her experience and recommendations about Down Syndrome individuals' employment. & By engaging in conversations with the persona, Jane aims to gather firsthand insights about employment strategies for individuals with Down Syndrome, which will inform and improve the effectiveness of her design solutions. \\ \hline 
         (3) Mark questions &  Jane needs a way to highlight important questions and dismiss those that are unnecessary, as she often generates many questions during her research. & Mark questions will enable Jane to streamline her focus, ensuring she prioritizes and addresses the most critical questions that will directly impact her design decisions. \\ \hline 
         (4) Historical chats&  Jane is not satisfied with the current chat content, she wants to start a new chat and still has access to previous chat records.& Historical chats will enable Jane to start fresh discussions as needed, without losing historical data, allowing her to compare new insights with past interactions and continuously refine her understanding and approach.\\ \hline 
        (5) Summarize questions*  &  Jane wants to compile an interview script by summarizing and organizing her questions effectively.& Summarizing questions will help ensure that Jane's interviews are structured and focused, allowing her to gather comprehensive and relevant information efficiently during her interactions with the LLM persona. \\ \hline 
        (6) Detailed Interview scripts&  Jane needs the ability to view all details of summarized interview scripts to ensure she has a comprehensive understanding of the content covered.& Detailed Interview scripts will allow Jane to make informed decisions, refine her approach, and ensure that no critical information is missed during her interactions and analyses.\\ \hline
         (7) Persona Library&  Jane needs a way to save multiple personas she finds interesting so that she can revisit and refine them later, even though she can only process one persona at a time.& Persona library will allow Jane to build and maintain a versatile library of personas, enabling her to draw from a diverse set of insights and perspectives when needed for future projects.\\ \hline 
         
    \end{tabular}
    \caption{Interaction Phase - Conversational System Features. *Summarize Questions feature was not operational at the time of the study, but it was shown in the UI}
    \label{tab:features}
\end{table}
\end{document}